\documentclass{article}

\usepackage{tablefootnote}

\usepackage{arxiv}
\usepackage[utf8]{inputenc} 
\usepackage[T1]{fontenc}    
\usepackage{hyperref}       
\usepackage{url}            
\usepackage{booktabs}       
\usepackage{amsfonts}       
\usepackage{nicefrac}       
\usepackage{microtype}      
\usepackage{lipsum}
\usepackage{graphicx}
\graphicspath{ {images/} }
\usepackage{enumitem}

\usepackage{subcaption}
\usepackage{float}
\usepackage{mathtools}

\usepackage{tikz}
\usepackage{tikz-dependency}
\usetikzlibrary{fit,positioning,arrows.meta,shapes}
\tikzset{neuron/.style={shape=circle, minimum size=1.25cm, 
  inner sep=0, draw, font=\small}, io/.style={neuron, fill=gray!20}}
\usetikzlibrary{shapes,arrows}
\usetikzlibrary{calc}
\usetikzlibrary{positioning}

\usepackage[symbol]{footmisc}

\title{Deep reinforcement learning on a multi-asset environment for trading}

\author{
 Ali Hirsa \\
  IEOR Department\\
  Data Science Institute \\
  Columbia University\\
  \texttt{ali.hirsa@columbia.edu} \\
   \And
 Joerg Osterrieder\footnotemark[1] \\
  School of Engineering\\
  Zurich University of Applied Sciences\\
  Winterthur, Switzerland \\
  \texttt{joerg.osterrieder@zhaw.ch} \\
  \\
The Hightech Business and Entrepreneurship Group\\Faculty of Behavioural, Management and Social Sciences\\University of Twente\\Enschede, Netherlands
\\
  \texttt{joerg.osterrieder@utwente.nl}

   \And
 Branka Hadji Misheva\footnotemark[1] \\
  School of Engineering\\
  Zurich University of Applied Sciences\\
  Winterthur, Switzerland \\
  \texttt{branka.hadjimisheva@zhaw.ch} \\
   \And
  Jan-Alexander Posth\footnotemark[1] \\
  Institute of Wealth \& Asset Management\\
  ZHAW School of Management and Law\\
  CH-8401 Winterthur, Switzerland\\
  \texttt{posh@zhaw.ch} \\
}

\begin{document}
\maketitle
\footnotetext[1]{
Financial support by the Swiss National Science Foundation within the project “Mathematics and Fintech - the next revolution in the digital transformation of the Finance industry” is gratefully acknowledged by the corresponding author. 
This research has also received funding from the European Union's Horizon 2020 research and innovation program FIN-TECH: A Financial supervision and Technology compliance training programme under the grant agreement No 825215 (Topic: ICT-35-2018, Type of action: CSA) and from Innosuisse under the grant agreement "Strengthening Swiss Financial SMEs through Applicable Reinforcement Learning" (Innosuisse Innovation project 47959.1 IP-SBM).
Furthermore, this article is based upon work from the COST Action 19130 Fintech and Artificial Intelligence in Finance, supported by COST (European Cooperation in Science and Technology), www.cost.eu (Action Chair: Joerg Osterrieder).\newline
The authors are grateful to Stephan Sturm, Matthew Dixon, Paul Bilokon, Saeed Amen, Chris Auth, Aasma John, Shengnan Cen, Wunan Hou, Xinyi Shu,
management committee members of the COST (Cooperation in Science and Technology) Action Fintech and Artificial Intelligence in Finance as well as speakers and participants of the $4^{\text{th}}$ and $5^{\text{th}}$ European COST Conference on Artificial Intelligence in Finance and Industry, which took place at Zurich University of Applied Sciences, Switzerland, in September 2019 and 2020.}

\begin{abstract}
Financial trading has been widely analyzed for decades with market participants and academics always looking for advanced methods to improve trading performance. Deep reinforcement learning (DRL), a recently reinvigorated method with significant success in multiple domains, still has to show its benefit in the financial markets. We use a deep Q-network (DQN) to design long-short trading strategies for futures contracts. The state space consists of volatility-normalized daily returns, with buying or selling being the reinforcement learning action and the total reward defined as the cumulative profits from our actions. Our trading strategy is trained and tested both on real and simulated price series and we compare the results with an index benchmark. We analyze how training based on a combination of artificial data and actual price series can be successfully deployed in real markets.
The trained reinforcement learning agent is applied to trading the E-mini S\&P 500 continuous futures contract. Our results in this study are preliminary and need further improvement.

\end{abstract}

\section{Introduction}

Reinforcement learning (RL) is an area of machine learning concerned with how agents ought to take actions in an environment to maximize the notion of cumulative reward. Traditionally, RL has been applied to the playing of several Atari games but recently more applications of RL have come up especially in the field of finance where trading challenges can be formulated as a game in which an agent is designed to maximize a reward. The breadth of using RL for automated financial trading is a widely discussed topic and an online decision-making process that involves two critical steps:

\begin{enumerate}[nosep]
    \item market condition summarization and, 
    \item optimal action execution
\end{enumerate}

Since the financial market is one of the most dynamic and fluctuating entities, the decision-making process gets more convoluted and challenging due to the absence of supervised information. It is different from the conventional learning tasks as it necessitates the agent to learn and explore an unknown environment and simultaneously make correct decisions. This pursuit of exploring and self-adapting is where RL comes into play. 

An investment decision in RL is a stochastic control problem, usually a Markov Decision Process (MDP), where the trading strategies are learned directly from the interactions with the market. This eliminates the necessity of building forecasting models for returns or futures prices. The success of RL has been very extensive in practical applications such as Atari game playing, helicopter control and robot navigation, sometimes even outperforming humans. However, with financial markets more research is needed to decide if a trained RL model can successfully beat human traders \cite{deng2016deep}.

RL presents a unique opportunity to model the complexities of trading in which traditional supervised learning models may not be able to succeed. Future contracts trading is one such financial problem. Some common challenges \cite{deng2016deep} \cite{zhang2020deep} that arise while applying RL to financial trading are:

\begin{enumerate}[nosep]
    \item \textbf{Data and Its Characteristics:} Data in financial markets is non-Markovian, highly non-stationary, and has a low signal-to-noise ratio.
    \item \textbf{Exploration and Exploitation:} Financial markets are very dynamic and placing actual orders to investigate the benefit of new trading strategies can be quite expensive, costs depending on trading frequency, turnover, volatility, liquidity, and transaction fees (the latter two being not so much of an issue for futures contracts). In most cases, a previous simulation (backtest) of any new proposed strategy is needed in order to assess its risk profile and its viability. This, in turn, introduces the problem of over-fitting which needs to be dealt with in a statistically prudent way.
    \item \textbf{Lack of baseline:} RL has demonstrated huge success in the video game and robotics area, but so far, financial trading is lacking behind and thus there is no clear baseline nor a suitable MDP model, or a set of hyperparameters that can be used for reference \cite{huang2018financial}. 
\end{enumerate}

We adopt the classical DQN algorithm to trade E-mini S\&P 500 continuous, roll-adjusted futures contracts. Our algorithm is trained and tested using both real and simulated price series data and the results are compared the performance characteristics of an index benchmark, the S\&P 500 index itself. In order to extend the state space, time series data of different asset classes is explored, including commodities, equity indexes, fixed income, and FX markets. To account for the fact that real training data is rather limited, simulated data is constructed via a geometric Brownian motion (GBM) price process as well as a variance gamma price process (VG) with parameters estimated from real data. As also observed with the conventional optimization of trading strategies, our experiments show that the DQN algorithm is prone to over-fitting if not done carefully, due to the limited amount of training data compared to the size of the neural network parameters. For this rather generic challenge of any high-dimensional optimization, careful analysis and discussion of the results is needed.

The structure of this paper is as follows. Section 2 gives a brief description of the current literature. Section 3 presents the MDP setup and describes the DQN model we used. Section 4 explains the experiments, including our data, training schemes, and the results. Section 5 concludes and provides an outlook.

\section{Literature Review}

Based on different learning objectives and setup requirements, current literature on RL in trading can be categorized into three main types: critic-only, actor-only, and actor-critic approach. In this section, we review five research papers that use different approaches and methodologies to derive trading strategies. Detailed descriptions of those can be found in Appendix \ref{appendix literature}. For a comprehensive analysis of the applicability of self-play algorithms for financial markets, see \cite{posth2021RL}.

In the critic-only approach, discrete action spaces are used, a state-action value function, $Q$, is created to represent how good a specific action is in that space and an agent is trained to go long or short. While making the decision, the agent senses the state of the environment and chooses an action that presents the best outcome according to the value function $Q$. Among the critic-only algorithm, DQN is the most used framework in this field (Neuneier 1996 \cite{neuneier1996optimal}; Dempster et al. 2001; Dempster and Romahi 2002; Bertoluzzo and Corazza 2012; Jin and El-Saawy 2016; Ritter 2017; Huang 2018). Olivier Jin and Hamza El-Saawy \cite{jin2016portfolio} used DQN to manage a portfolio containing two stocks with different volatility. They take a set of discrete numbers as action spaces to represent the percentages of the portfolio’s total value each stock composes and modify the portfolio return by penalizing volatility as reward functions. By adopting an $\epsilon$ - greedy strategy, the agent is encouraged to randomly explore the state space. To stabilize the neural networks, they also add the experience replay to de-correlate the time series data. Noted that when using DQN to train the agent, each paper adopts a different methodology, mainly $Q$ function approximators, state spaces, action spaces, reward functions, etc. Zhengyao Jiang, Dixing Xu and Jinjun Liang \cite{jiang2017deep} implemented the RL framework by using three different underlying networks, a Convolutional Neural Network (CNN), a basic Recurrent Neural Network (RNN), and a Long Short-Term Memory (LSTM) to manage a cryptocurrencies portfolio. To dive deeper into the combination of RNN and DQN, Chien-Yi Huan \cite{huang2018financial} modified the existing deep recurrent Q-network algorithm (DRQN) to be suitable for the financial trading task by using a smaller replay memory and sampling a longer sequence for training. The model has achieved positive returns on twelve different currency pairs net of transaction costs.

The next common approach is the actor-only approach (Moody et al. 1998; Moody and Saffell 2001; Deng et al. 2016; Lim, Zohren, and Roberts 2019). The advantage of using this approach is that the learning process converges faster and the action space of the agent can be continuous. Here, the agent senses the state of the environment and acts directly without computing and comparing the expected outcomes of different actions. The agent hence learns a direct mapping from states to actions. 

The last one is the actor-critic approach. This approach combines the advantages of both the critic-only and the actor-only approach. The idea behind this approach is to use both an actor and a critic. Given the current state of the environment, the actor determines the action of the agent, and the critic judges the action that was selected. In this manner, the actor learns to select the action that is considered best by the critic and the critic learns to improve its judgment. This approach, however, is the least studied one in financial applications (Li et al. 2007; Bekiros 2010; Xiong et al. 2018) and has only a limited number of supporting works.

\section{Methodology}

In this section, we introduce our setup including state space, action space, and reward function. We also describe the DQN model, a DRL algorithm, which we train to trade the S\&P500 continuous, roll-adjusted futures contract.

\subsection{Markov Decision Process}
The trading problem is considered as an MDP which contains states, actions, a reward function, and a discount factor. The process starts from an agent receiving some information about the environment denoted as a state $S_0$. Based on that state, the agent chooses an action $A_0$ and a reward of $R_1$ is given to the agent at the next step. The agent then enters another new state $S_1$ and gets into another round of action, reward, and state. The loops of interaction between the agent and the environment produces a trajectory $\tau$ = [$S_0$, $A_0$, $R_1$, $S_1$, $A_1$, $R_2$, $S_2$, $A_2$, $R_3$, $\dots$]. The goal of reinforcement learning is to maximize the expected cumulative value of reward at any time t, which is denoted as $G_t$:
$$
G_t = E[\sum_{k=t+1}^{T} \gamma^{k-t-1} R_k]
$$
where $\gamma$ is the discount factor.

\paragraph{State Space:}
In order to account for the most important frequency regimes of returns, in our model we take returns $(r_t)$ over the past 1-day, 1-month, 2-month, 3-month,  6-month and 1-year periods to represent the states. The returns are normalized by the daily volatility scaled to the appropriate time scale following \cite{lim2019enhancing}. For instance, normalised annual returns are given by $r_{t-252, t}^{(i)} /\left(\sigma_{t}^{(i)} \sqrt{252}\right)$ where $\sigma_{t}^{(i)}$ is computed using an exponentially weighted moving standard deviation of $r_t^{(i)}$ with a 63-day (3-month) span. By this, our state space will be fully risk-adjusted, i.e. scaled to variations in volatility, with the focus on more recent volatility. At any given time step, we take the past 30 observations of each feature to form a single state.

\paragraph{Action Space:}
We only study discrete action spaces here for simplicity. An action set of $\{-1, 0, 1\}$ is used to represent the position directly, i.e. $1$ corresponds to a full long position, $-1$ to a full short position, and $0$ to no holdings. When the previous action and current action are different, a transaction cost composed of proportional cost and fixed cost will occur.

\paragraph{Reward Function:}
We define the reward function as the product of asset price percentage change and current action,
$$
R_t = A_t \left( \frac{p_{t+1}}{p_t}-1 \right).
$$
where is $A_t$ current action; $p_t$, $p_{t+1}$ the corresponding prices at the current and next time step.

\subsection{DQN Algorithm}\label{DQN algorithm}
The DQN algorithm allows the agent to explore the unstructured environment and acquire knowledge which makes it possible to map environment states to agent actions through an approximation of the state-action value function ($Q$ function). The objective is to derive the optimal nonlinear function $Q(S, A)$ with a neural network. Suppose the $Q$ function is parameterized by some $\theta$. To get the approximated $Q$ function, the algorithm keeps tuning function parameters by minimizing the loss function which is defined as the mean squared error between the current and target $Q$:
$$
L(\theta) = E[(Q_\theta(S,A)-Q'_\theta(S,A))^2]
$$
$$
Q'_\theta(S_t,A_t) = r+\gamma\,\text{argmax}_{A'}(S_{t+1},A_{t+1})
$$
where $L(\theta)$ is the loss function and $Q$ is the value function.

\section{Experiments}

We construct different trading strategies, using both simulated data and real data, and compare the results to the index. 

\subsection{Data Preparation}
Both real and simulated time-series data for the S\&P 500 futures contract and data from further 25 liquid futures contracts, from different asset classes, is used to train the RL agent with the algorithm being evaluated on trading the S\&P 500 futures E-mini contracts.

\subsubsection{Multi-asset futures contracts}\label{Multi-asset futures contracts}
We use time-series data on 25 continuous front-month futures contracts with roll methodology OR (backwards ratio, open interest switch). The data ranges from 06/2000 to 05/2019 and consists of futures prices for commodities, equity indices, fixed income, and FX. A detailed description of each contract can be found in Appendix \ref{appendix data}, Table \ref{tab:contratcs}.

Our trading target is the E-mini S\&P 500 continuous futures contract. For the naive strategy, we use the volatility-normalized returns of the contract as the state space. As mentioned previously (see Section \ref{DQN algorithm}), we take returns $(r_t)$ over six different periods, and take the past 30 observations to form a single state. This results in a dimension of our observation space of $30 \times 6$. We train our model using five years worth of data and fix the model parameters for the next five years to produce out-of-sample results. We then apply a rolling forecast over the entire time period and obtain a combined out-of-sample performance.

To construct a more robust strategy, we use volatility-normalized returns of 25 futures contracts as state space - thus, the dimension is $30 \times 6 \times 25$. We then retrain our model on a quarterly (semi-annually/annually) basis, using all data available up to that point to optimize the parameters. Model parameters are fixed for the next quarter (semi-year/year) to produce out-of-sample results.

Here, clearly the motivation for extending the state space is the need for more data on which we can train the RL agent. By including 25 time series of futures contracts across four asset classes, we hope to a) achieve more robustness by reducing over-fitting, and to b) account for a broader range of possible market constellations, thus optimizing the adaptiveness of our agent to a rapidly changing environment. Furthermore, given the well-known stylized fact that inter-asset-class correlations between risk-on and risk-off assets are decidedly negative while, for example, intra-asset-class correlations in equities are mostly positive (\cite{papenbrock_handling_2015}), we expect the RL agent to infer from this additional information that is hidden in the interrelation structure of the various markets/assets. We are aware, of course, that the inclusion of non-equity related data into our state space poses the difficulty of fundamentally different return statistics exhibited by the different asset classes. However, we think that the trade-off with regard to learning - for the reasons mentioned above - is a positive one.

Furthermore, from an information-theoretic point of view as well as seen through the lenses of the efficient market hypothesis, we are interested in understanding if equity markets are already fully efficient or if additional asset classes are needed for a full understanding of the US equity market.

\subsubsection{Simulated price series}
The main purpose of using simulated time-series data is to supplement the real-world data, thus broadening the state space and increasing the robustness of the RL agent. We implement two different stochastic processes to simulate the price-series data of the S\&P 500 futures contract:

\paragraph{Geometric Brownian Motion:} 
The GBM is governed by the stochastic differential equation (SDE)
$$
d S_{t}=\mu S_{t} d t+\sigma S_{t} d W(t)
$$ 
where $\mu$ is the drift and $\sigma$ is the volatility. 
Via Ito's lemma, the price process is calculated as 
$$
S_{t}=S_{0} \exp \left\{\left(\mu-\frac{\sigma^{2}}{2}\right)t +\sigma \sqrt{t} z\right\}
$$ 
where $S_0$ is the price at time 0, $s_t$ is the price at time t, and $z \sim \mathcal{N}(0,1)$.

\paragraph{Variance Gamma:} 
The VG process $X(t ; \sigma, \nu, \theta)$ is obtained by evaluating a Brownian motion with drift $\theta$ and volatility $\sigma$ at a random time given by a gamma process $\gamma(t ; 1, \nu)$, with mean rate unity and variance rate $\nu$ as
$$
X(t ; \sigma, \nu, \theta)= \theta \gamma(t ; 1, \nu) + \sigma W(\gamma(t ; 1, \nu))
$$
where $\theta$ and $\nu$ controls skewness and kurtosis. The price process is calculated as
$$
S_{t}=S_{0} \exp\left\{(\mu + \omega) t+X(t ; \sigma, \nu)\right\}
$$ 
where $\omega=\frac{1}{\nu} \ln \left(1-\theta \nu - \sigma^{2} \nu / 2\right)$. 

The VG process allows for a wider modeling of skewness and kurtosis than the Brownian motion does and hence is expected to better account for tail events in the probability distribution of returns. This is an important characteristic of the VG process as it is widely acknowledged that equity returns do not follow a normal distribution but exhibit "fat tails", i.e. large losses occurring with a higher probability than suggested by the normal distribution assumption.

\begin{figure}[h!]
\begin{center}
\begin{tikzpicture}[scale = 2.0]
\node[circle, thick, fill=blue!20, minimum width=2.0cm, draw](1) at (2.0,0){up trend};
\node[circle, thick, fill=blue!20, minimum width=2.0cm, draw](2) at (4.0,0){no trend};
\node[circle, thick, fill=blue!20, minimum width=2.0cm, draw](3) at (6.0,0){down trend};
\draw[solid, thick, <-](2.0+0.5*0.707,0.5*0.707) to [out=35,in=145] (4.0-0.5*0.707,0.5*0.707);
\draw[solid, thick, ->](4.0+0.5*0.707,0.5*0.707) to [out=35,in=145] (6.0-0.5*0.707,0.5*0.707);
\draw[solid, thick, ->](2.0+0.5*0.707,-0.5*0.707) to [out=-35,in=-145] (4.0-0.5*0.707,-0.5*0.707);
\draw[solid, thick, <-](4.0+0.5*0.707,-0.5*0.707) to [out=-35,in=-145] (6.0-0.5*0.707,-0.5*0.707);
\draw[solid, thick, <-](2+0.5*0.5735,0.5*0.8192    ) to [out=45, in=-45] (2+0.5*0.5735,0.5*0.8192+0.5);
\draw[solid, thick, - ](2+0.5*0.5735,0.5*0.8192+0.5) to [out=135,in=45 ] (2-0.5*0.5735,0.5*0.8192+0.5);
\draw[solid, thick, - ](2-0.5*0.5735,0.5*0.8192+0.5) to [out=-135,in=135](2-0.5*0.5735,0.5*0.8192    );
\draw[solid, thick, <-](4+0.5*0.5735,0.5*0.8192    ) to [out=45, in=-45] (4+0.5*0.5735,0.5*0.8192+0.5);
\draw[solid, thick, - ](4+0.5*0.5735,0.5*0.8192+0.5) to [out=135,in=45 ] (4-0.5*0.5735,0.5*0.8192+0.5);
\draw[solid, thick, - ](4-0.5*0.5735,0.5*0.8192+0.5) to [out=-135,in=135](4-0.5*0.5735,0.5*0.8192    );
\draw[solid, thick, <-](6+0.5*0.5735,0.5*0.8192    ) to [out=45, in=-45] (6+0.5*0.5735,0.5*0.8192+0.5);
\draw[solid, thick, - ](6+0.5*0.5735,0.5*0.8192+0.5) to [out=135,in=45 ] (6-0.5*0.5735,0.5*0.8192+0.5);
\draw[solid, thick, - ](6-0.5*0.5735,0.5*0.8192+0.5) to [out=-135,in=135](6-0.5*0.5735,0.5*0.8192    );
\node at (2,1.2) {\footnotesize $P_{UU}$};
\node at (4,1.2) {\footnotesize $P_{NN}$};
\node at (6,1.2) {\footnotesize $P_{DD}$};
\node at (3.0,0.75) {\footnotesize $1\!-\!\frac{P_{NN}}{2}$};
\node at (5.0,0.75) {\footnotesize $1\!-\!\frac{P_{NN}}{2}$};
\node at (3.0,-0.75) {\footnotesize $1\!-\!P_{UU}$};
\node at (5.0,-0.75) {\footnotesize $1\!-\!P_{DD}$};
\end{tikzpicture}
\end{center}
\caption{3-state problem (see \cite{ali2020deep})}
\label{fig:3coin}
\end{figure}
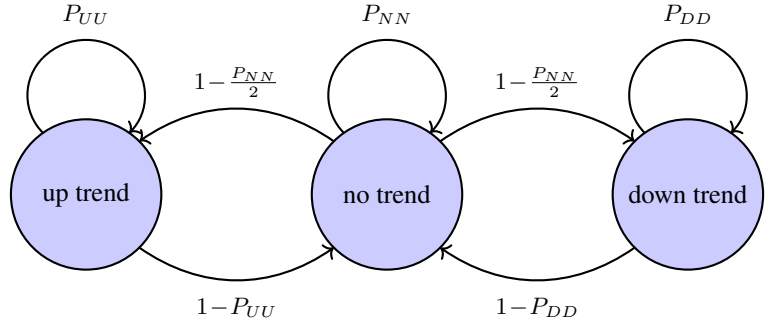

Furthermore, the stock prices in equity markets are know to transition between different volatility regimes (see e.g., \cite{papenbrock_handling_2015}, \cite{alex206773}): Depending on whether one is experiencing an up-, down-, or sideways-trend, volatility can be vastly different, resulting in different return statistics for the given trend period. To account for these different volatility regimes, we select different time intervals of the E-mini S\&P 500 continuous futures contract data to estimate three different sets of model parameters, i.e up-trend (low volatility), no-trend (medium volatility), and down-trend (high volatility) as shown in Table \ref{tab:para}. Then we introduce a 3-state problem with transitions among these three states. The state transition diagram of the 3-state problem is drawn in Figure \ref{fig:3coin}. By this we ensure that our artificially generated training data reflects all of the most important states of the price time-series in the given financial market as well as the appropriate switching behaviour between these states. The selected intervals of real time-series data for the parameter estimation and examples of the GBM and the VG processes are shown in Appendix \ref{appendix data}, Figure \ref{fig:regimes} (left and right, respectively).

\begin{table}
\caption{Estimated annualized parameters for the three trend regimes}
\centering
\begin{tabular}{lrrr|lrrr}
\hline
\multicolumn{4}{c|}{GBM} & \multicolumn{4}{c}{VG} \\\hline
 & \multicolumn{1}{c}{up} & \multicolumn{1}{c}{no} & \multicolumn{1}{c|}{down} &  & \multicolumn{1}{c}{up} & \multicolumn{1}{c}{no} & \multicolumn{1}{c}{down} \\\hline
$p_{ii}$& 0.950 & 0.900 & 0.950 & $p_{ii}$& 0.950 & 0.900 & 0.950 \\
$S_0$ & 1051.344 & 1051.344 & 1051.344 & $S_0$ & 1051.344 & 1051.344 & 1051.344 \\
drift ($\mu$)& 0.254 & 0.016 & -0.440 &drift ($\mu$) & 0.254 & 0.016 & -0.440 \\
volatility ($\sigma$) & 0.109 & 0.158 & 0.441 & volatility ($\sigma$) & 0.109 & 0.158 & 0.441 \\
 &  &  &  & skewness ($\theta$) & -0.742 & -0.287 & -0.410 \\
 &  &  &  & kurtosis ($\nu$) & 3.93e-04 & 2.44e-04 & 2.74e-04\\ \hline
\end{tabular}
\label{tab:para}
\end{table}


As for the naive strategy with real time-series data, we get an observation space of dimension $30 \times 6$ and train our model using the equivalent of five years of simulated data. Model parameters are fixed for the next five years to produce out-of-sample results.

\subsection{Training Schemes}
In our work, we use the stable baselines DQN model \cite{stable} with an \textit{LnMlpPolicy}. We implement the policy using a multi-layer perceptron (MLP) with two hidden layers, with 64 units, with $tanh$ as activation function, and with layer normalization. 

\subsubsection{Model architecture}
The architecture of the neural network for our naive DQN model is listed in Table \ref{tab:arch} and the shape of parameters for each layer is shown in Appendix \ref{appendix model}, Table \ref{tab:modparas}. The network is relatively small with 64,521 parameters. For the model using the returns of 25 futures contracts as a state, the architecture is the same except the units of the input layer increases to 4,500, and the total amount of parameters increases to 1,170,441. 

As mentioned previously (see Section \ref{Multi-asset futures contracts}), we use five years of data to train the model, so the training data size is approximately 1,200. It is therefore rather limited, especially for the naive model. To increase the sample size, we include simulated data for training.

\begin{table}
\centering
\caption{Model architecture for naive DQN model}
\begin{tabular}{l|lr|l|lr}
\hline
Function & Layer & Units & Function & Layer & Units \\\hline
Action-Value & Input Layer & 180 & State-Value & Input Layer & 180 \\
 & Layer Norm &  &  & Layer Norm &  \\
 & Hidden Layer 1 & 64 &  & Hidden Layer 1 & 64 \\
 & Layer Norm &  &  & Layer Norm &  \\
 & Hidden Layer 2 & 64 &  & Hidden Layer 2 & 64 \\
 & Output Layer & 3 &  & Output Layer & 1\\\hline
\end{tabular}
\label{tab:arch}
\end{table}

\subsubsection{Hyperparameters}
The Adam optimizer is used for training networks, and the following hyperparameters are tuned via build-in parameters of the stable baselines DQN model:

\begin{itemize}
    \itemsep0em
    \item learning\_rate: learning rate for Adam optimizer
    \item gamma: discount factor
    \item batch\_size: size of a batched sampled from replay buffer for training
    \item buffer\_size: size of the replay buffer
    \item exploration\_fraction: fraction of entire training period over which the exploration rate is annealed
    \item learning\_starts: how many steps of the model to collect transitions for before learning starts
    \item target\_network\_update\_freq: update the target network every target\_network\_update\_freq steps
    \item prioritized\_replay: use prioritized replay buffer or not.
\end{itemize}

We validate our approach on the S\&P 500 E-mini futures contract, the values of the hyperparameters are listed in Table \ref{tab:hyper} and are kept fixed across all experiments.

\begin{table}
\caption{Values of hyperparameters}
\centering
\begin{tabular}{l|r}
\toprule
    learning\_rate & 0.001 \\
    gamma & 0.94 \\
    batch\_size & 128 \\
    buffer\_size & 30000 \\
    exploration\_fraction & 0.25 \\
    learning\_starts & 100 \\
    target\_network\_update\_freq & 500\\
    prioritized\_replay & True\\
\bottomrule
\end{tabular}
\label{tab:hyper}
\end{table}

\subsection{Experimental Results}
We compare the performance of our model with the long-only index and we calculate the cumulative returns either gross or net of transaction costs for each strategy. Following \cite{zhang2020deep}, we evaluate the performance using the following metrics (for the sake of simplicity and without restricting generality, we assume the risk-free rate to be $0\%$):

\begin{itemize}
    \itemsep0em
    \item $\text{E(R)}$: annualized expectation of daily return
    \item $\text{std(R)}$: annualized standard deviation of daily return
    \item Downside Deviation ($\text{DD}$): annualized standard deviation of negative trade returns, also known as downside risk
    \item Sharpe Ratio: annualized Sharpe Ratio ($\text{E(R)}/\text{std(R)}$)
    \item Sortino Ratio: a variant of Sharpe Ratio that uses downside deviation as risk measures ($\text{E(R)}/\text{DD}$)
    \item Maximum Drawdown ($\text{MDD}$): shows the maximum observed loss from any peak of a portfolio
    \item Calmar Ratio: compares $\text{E(R)}$ with $\text{MDD}$. In general, the higher the ratio is, the better the performance is
    \item $\%$ +ve Returns: percentage of positive daily returns
    \item $\frac{\text{Ave. P}}{\text{Ave. L}}$: the ratio between the number of observations of positive and negative daily returns.
\end{itemize}

\subsubsection{Naive strategies}
We first build the models with 1 / 50 / 90 paths of a simulated process as training data. The histograms of the annualized Sharpe ratio of 100 out-of-sample predictions are shown in Appendix \ref{appendix result}, Figure \ref{fig:gbm} and \ref{fig:vg}. There are three important observations we can make: First, as the number of paths increases, the average out-of-sample performance improves. Second, for the no-trend regime the results are rather ambiguous - as should be expected. This can be explained by the fact that the no-trend regime does not contain valid directional information, so the expectation of returns for both model (no trend signal for trading) and benchmark (by definition of the regime) should be approximately zero. Third, compared to the benchmark, there seems to be a significant out-performance of the model for the down-trend regime. A possible reason for this statistical behaviour could be that our model allows for short positioning while the benchmark does not.

\begin{figure}[h!]
     \centering
     \begin{subfigure}[b]{0.33\textwidth}
         \centering
         \includegraphics[width=\textwidth]{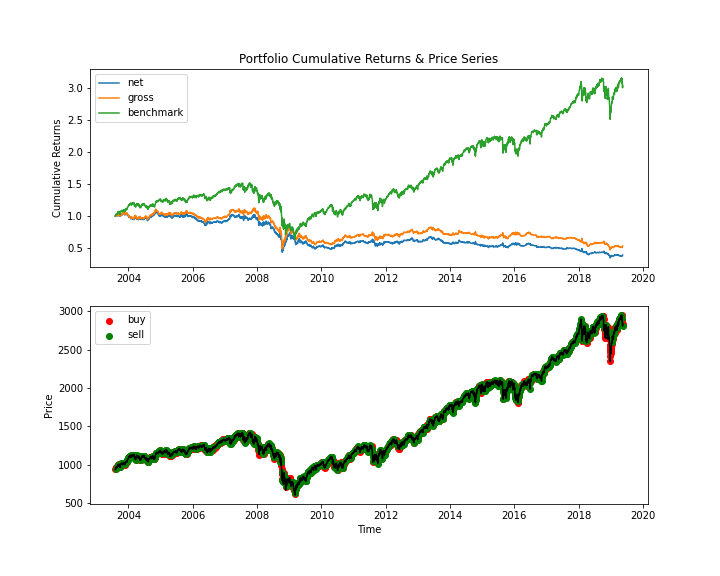}
         \caption{Train on rolling 5-year real data}
     \end{subfigure}
     \hfill
     \begin{subfigure}[b]{0.33\textwidth}
         \centering
         \includegraphics[width=\textwidth]{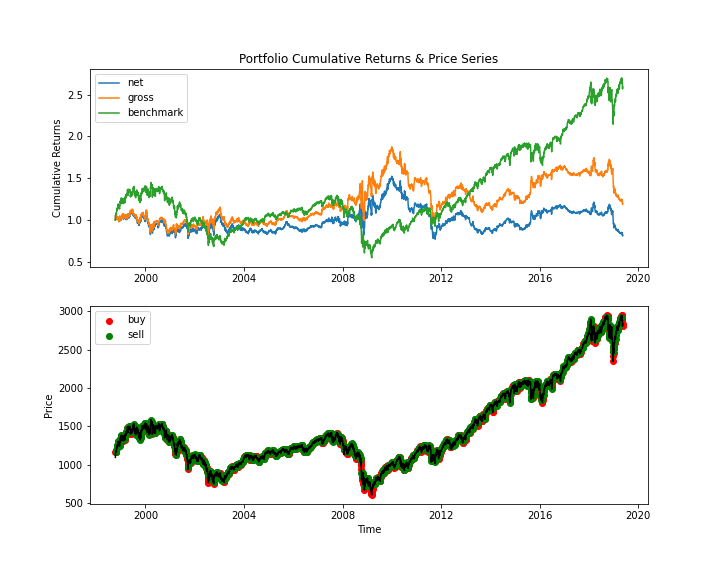}
         \caption{Train on GBM process}
     \end{subfigure}
     \hfill
     \begin{subfigure}[b]{0.33\textwidth}
         \centering
         \includegraphics[width=\textwidth]{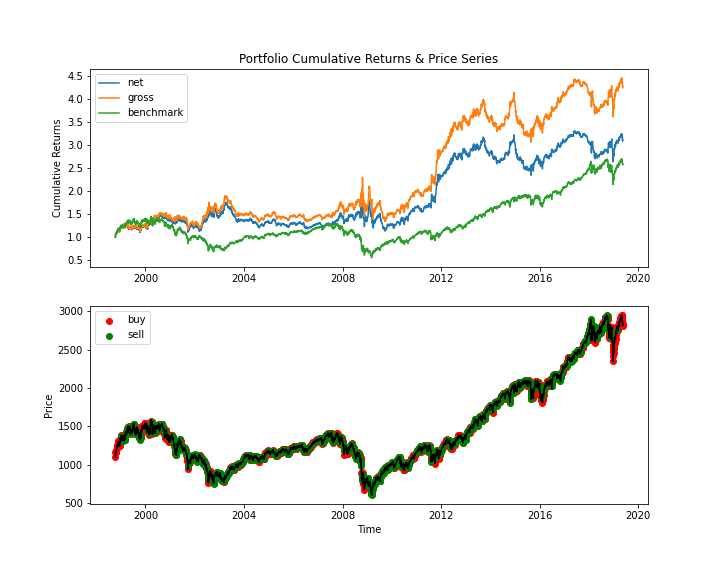}
         \caption{Train on VG process}
     \end{subfigure}
    \caption{Cumulative returns of naive strategies with different training data sets}
    \label{fig:naive}
\end{figure}

The cumulative returns of naive strategies with real and simulated training data sets along with their respective trading signals are presented in Figure \ref{fig:naive}. We also present performance metrics in Table \ref{tab:naive}. The table is split into three parts based on the three types of different training data: (a) represents rolling 5-year actual time-series data, (b) the GBM process, and (c) the VG process. The results show the performance of the model with or without transaction costs, and we compare them with the index benchmark's performance. We observe that model (c) trained with 50 paths of the VG process delivers better performance than the benchmark (net: $7.3\%$, gross: $9.0\%$ vs. benchmark: $6.6\%$), while the other two models fail to outperform the benchmark (net: $-4.4\%$ / $1.0\%$, gross: $-2.6\%$ / $2.4\%$ vs. benchmark: $9.0\%$ / $6.6\%$, for (a) and (b) respectively). The same holds true on a risk-adjusted basis, as measured by the Sharpe ratio. We note that even for the out-performing strategy (c), the ratio between the number of observations of positive and negative daily returns, $\frac{\text{Ave. P}}{\text{Ave. L}}$, is smaller (below one) than for the benchmark. This hints at the typical trading behaviour of an AI agent, i.e. making quite a high number of wrong trading calls with rather limited negative impact and a smaller number of right calls, albeit with substantial positive impact.

\begin{figure}[h!]
     \centering
     \begin{subfigure}[b]{0.66\textwidth}
         \centering
         \includegraphics[width=\textwidth]{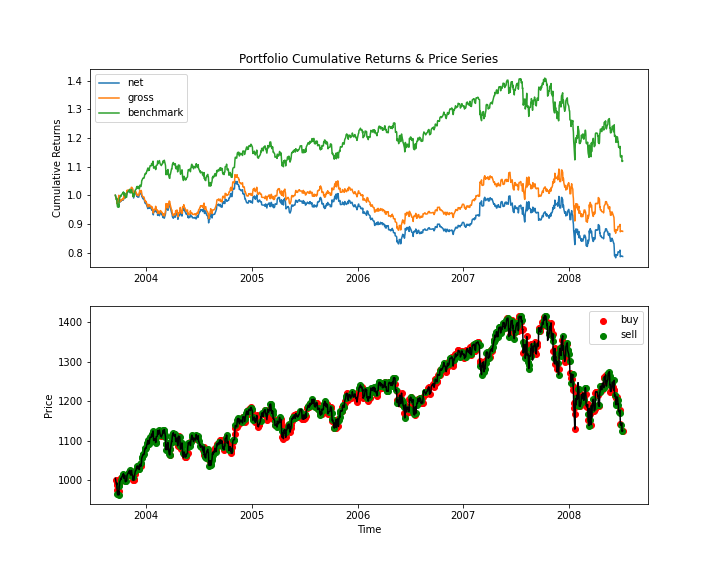}
     \end{subfigure}
    \caption{Out-of-sample cumulative returns of the naive strategy (a), for the first five years}
    \label{fig:overfitting}
\end{figure}

Finally, we present a close-up of the out-of-sample performance of the naive model (a) for the first five years in Figure \ref{fig:overfitting}.

\begin{table}
\centering
\caption{Experimental results for the naive strategies}
\begin{tabular}{l|l|rrrrrrrrr}
\hline
\multicolumn{2}{c|}{MODEL} & $\text{E(R)}$ & $\text{std(R)}$ & $\text{DD}$ & Sharpe & Sortino & $\text{MDD}$ & CALMAR & $\%$ +ve Rets & $\frac{\text{Ave. P}}{\text{Ave. L}}$ \\ \hline
(a) & net & -0.044 & 0.163 & 0.131 & -0.270 & -0.338 & 0.677 & -0.065 & 0.400 & 0.666 \\
 & gross & -0.026 & 0.164 & 0.139 & -0.157 & -0.184 & 0.581 & -0.044 & 0.402 & 0.672 \\
 & benchmark & 0.090 & 0.181 & 0.149 & 0.495 & 0.601 & 0.575 & 0.156 & 0.544 & 1.195 \\\hline
(b) & net & 0.005 & 0.171 & 0.130 & 0.030 & 0.040 & 0.494 & 0.011 & 0.405 & 0.681 \\
 & gross & 0.024 & 0.171 & 0.136 & 0.139 & 0.175 & 0.477 & 0.050 & 0.408 & 0.690 \\
 & benchmark & 0.066 & 0.190 & 0.147 & 0.346 & 0.448 & 0.621 & 0.106 & 0.532 & 1.139 \\\hline
(c) & net & 0.073 & 0.179 & 0.137 & 0.409 & 0.533 & 0.422 & 0.173 & 0.422 & 0.729 \\
 & gross & 0.090 & 0.179 & 0.145 & 0.501 & 0.620 & 0.416 & 0.216 & 0.423 & 0.734 \\
 & benchmark & 0.066 & 0.190 & 0.147 & 0.346 & 0.448 & 0.621 & 0.106 & 0.532 & 1.139 \\ \hline
\end{tabular}
\label{tab:naive}
\end{table}

\subsubsection{Advanced strategies}
We use a set of 25 liquid, multi-asset futures to train our agent and call the resulting strategy advanced. The strategy is evaluated on a rolling basis, being updated on a quarterly (1st), semi-annually (2nd) and annually (3rd) rolling basis.

\begin{figure}[h!]
     \centering
     \begin{subfigure}[b]{0.33\textwidth}
         \centering
         \includegraphics[width=\textwidth]{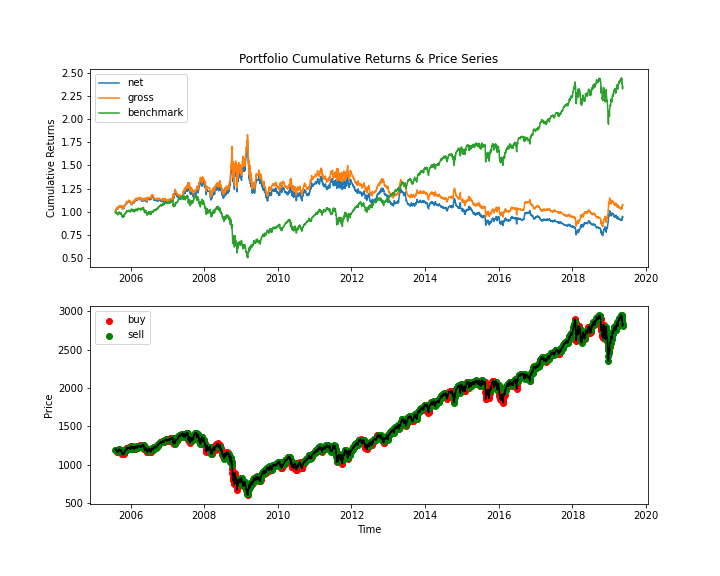}
         \caption{Quarterly}
     \end{subfigure}
     \hfill
     \begin{subfigure}[b]{0.33\textwidth}
         \centering
         \includegraphics[width=\textwidth]{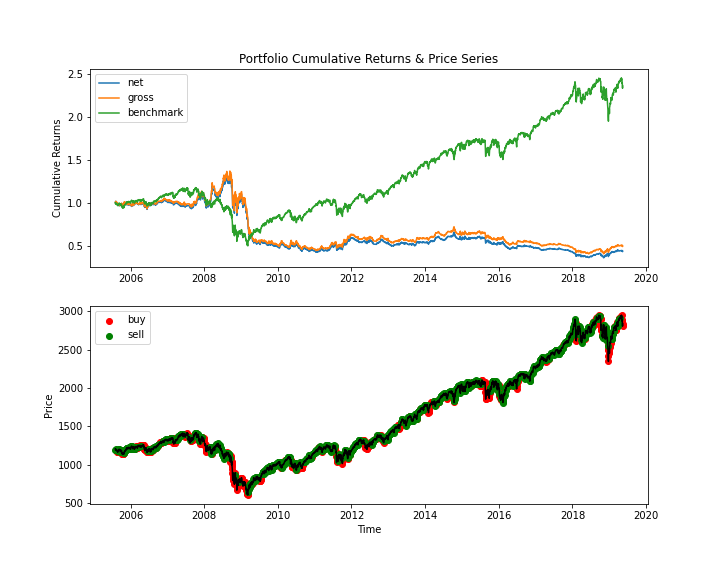}
         \caption{Semi-annually}
     \end{subfigure}
     \hfill
     \begin{subfigure}[b]{0.33\textwidth}
         \centering
         \includegraphics[width=\textwidth]{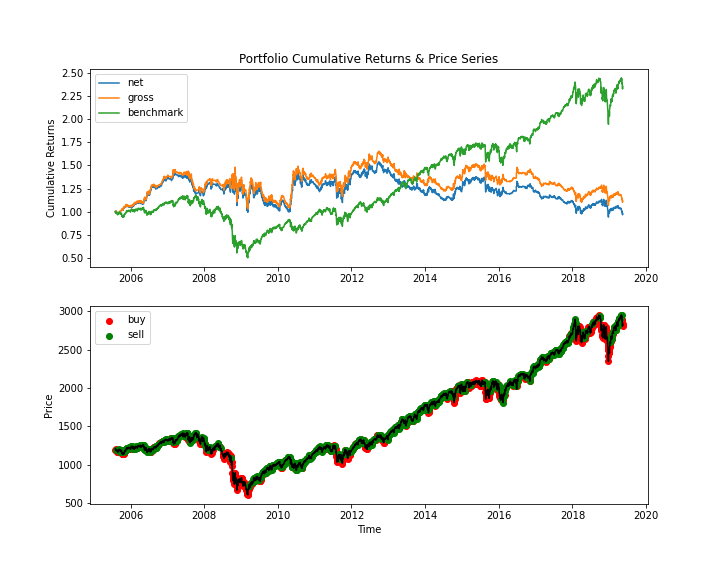}
         \caption{Annually}
     \end{subfigure}
    \caption{Cumulative returns of advanced strategies, with 25 futures as training data, using all data available up to the test day to train, updating on a quarterly (1st), semi-annually (2nd) and annually (3rd) rolling basis}
    \label{fig:real}
\end{figure}

The cumulative returns of those three advanced strategies are presented in Figure \ref{fig:real}, and the related performance metrics are in Table \ref{tab:advance}. None of the advanced strategies manages to beat the long-only benchmark. This is noteworthy since a) the strategy trained on the artificial data generated via the VG process exhibits an out-performance (net: $7.3\%$, gross: $9.0\%$) and b) two of the three strategies (quarterly and annually) at least manage to avoid the 2008 downturn and thus seem to offer consistent downside protection. Since after 2008 the real data has a distinct positive trend, arguably the most reasonable strategy in this case might be just to hold a long position \cite{zhang2020deep}. But the question remains why the RL agent in the case of the up-trend market was not able to adapt to this during its learning process and just enter a long position after 2008. This, together with the avoidance of the downturn in 2008, would have resulted in a very attractive overall out-performance. Finally, we note that the ratio between the number of observations of positive and negative daily returns, $\frac{\text{Ave. P}}{\text{Ave. L}}$, is comparable to the naive strategies. This implies that the RL agent, for all of the three advanced strategies, has not been able to infer a downside-protection strategy while retaining some upside.

\begin{table}
\centering
\caption{Experimental results for the advanced strategies}
\begin{tabular}{l|l|rrrrrrrrr}
\hline
\multicolumn{2}{c|}{MODEL} & $\text{E(R)}$ & $\text{std(R)}$ & $\text{DD}$ & Sharpe & Sortino & $\text{MDD}$ & Calmar & $\%$ +ve Rets & $\frac{\text{Ave. P}}{\text{Ave. L}}$ \\ \hline
(a) & net & 0.013 & 0.187 & 0.146 & 0.069 & 0.089 & 0.582 & 0.022 & 0.440 & 0.787 \\
 & gross & 0.023 & 0.187 & 0.149 & 0.120 & 0.151 & 0.543 & 0.041 & 0.442 & 0.792 \\
 & benchmark & 0.083 & 0.189 & 0.159 & 0.439 & 0.522 & 0.575 & 0.1446 & 0.533 & 1.142 \\\hline
(b) & net & -0.042 & 0.188 & 0.156 & -0.221 & -0.268 & 0.728 & -0.057 & 0.435 & 0.770 \\
 & gross & -0.033 & 0.188 & 0.160 & -0.173 & -0.205 & 0.703 & -0.047 & 0.436 & 0.772 \\
 & benchmark & 0.084 & 0.190 & 0.159 & 0.441 & 0.526 & 0.575 & 0.146 & 0.537 & 1.161 \\\hline
(c) & net & 0.014 & 0.178 & 0.139 & 0.079 & 0.102 & 0.388 & 0.036 & 0.401 & 0.669 \\
 & gross & 0.024 & 0.178 & 0.143 & 0.135 & 0.168 & 0.354 & 0.068 & 0.402 & 0.674 \\
 & benchmark & 0.084 & 0.191 & 0.159 & 0.442 & 0.529 & 0.575 & 0.146 & 0.540 & 1.172 \\\hline
\end{tabular}
\label{tab:advance}
\end{table}

\section{Conclusion and Outlook}

Our contribution is as follows:
\begin{enumerate}[nosep]
   \item Based on our results, we can state that is of great importance to chose the right generator for (artificial) financial market data to train the RL agent. Specifically, just simulating any time series data (e.g. with the GBM process) or just sourcing arbitrary historical data (e.g. via different markets and/or assets) is in the best case inefficient and in the worst case unproductive.
   \item Making the appropriate choice for the generating process of the training data, on the other hand, may lead to positive results and an out-performance of the RL strategy compared to the benchmark, even on a risk-adjusted basis. It seems vital that, next to the intelligent implementation of the actual AI agent training, the construction of the training data itself needs to be supported by insights into the time-series behaviour of financial markets and experience in statistical modelling.
\end{enumerate}

We adopt the DRL framework and the stable baselines DQN model to build long-short trading strategies for the S\&P 500 continuous futures contract. We choose a long-only strategy as the benchmark and compare its performance against our model's performance when using both simulated and real futures data. To mitigate the shortage of training data, multi-paths of simulated processes are used, which solves the over-fitting problem to some extent and improves the out-of-sample performance. We also construct advanced strategies by increasing the dimension of the state space and retraining the model on a rolling basis. 

For the naive strategies, we attribute the out-performance of the VG-trained model to three advantages of this approach: a) By being able to go short, the strategy, in principle, can profit during downturns - which the benchmark cannot. b) The VG model much better accounts for "fat tails" and hence should be able to deal with outliers and high-volatility regimes better than the other two approaches. c) The VG process, by construction, allows for more stable upside capture.
So while the avoidance of e.g. the 2008 downturn in the markets by the GBM model seems rather to be an artefact of the strategy being able to go short, the VG model rather consistently avoids downside volatility while at the same time preserving substantial upside. We are obviously aware a significant part of the positive performance of the VG model is generated during the years 2012, 2013, and 2014. This peculiarity warrants further statistical analysis since it, together with the downturn-avoidance and the relatively few "right" trading decisions, is suggestive of over-fitting.

For the advanced strategies, we believe the reasons for the bad performance of the RL agent are manifold: First, even with $25$ additional futures contracts the training data size is still rather limited compared with the model's parameter number. This, in principle, is the same challenge we have faced with the artificially generated data - only it is worse in this case because of even less data. Second, while the learning data generated via the VG process might have provided a quite good training proxy, i.e. retaining the vital return characteristics of the underlying market, the set of 25 multi-asset futures contracts does not seem to do so. Specifically, in this case the RL agent is not able to deduce additional valuable information from the other markets, asset classes, and their inherent correlation dependencies. Again, the dimensionality of the RL state space probably is much too small for the agent to learn an highly adaptive trading strategy, fit to deal with grossly different volatility regimes. Here, "more of the same" - as simulated in the VG training data - seems to be a better approach to the problem at hand and the challenge remains how to activate the RL agent to capitalize on the extra information inherent in the 25 multi-asset futures contracts - which certainly is there. Third, it remains to be analysed why the  evaluation frequency plays a vital role in the avoidance of the 2008 downturn, semi-annually being the negative exception here. This, to some extend, seems arbitrary. And finally fourth, the only strategy that managed to learn to trade with a positive result during an upward-trending market is the RL agent trained on VG-process generated data.

Concluding, we can state that the higher-order moments of the (simulated) return distribution on which the agent is trained are very important for capturing the correct statistical behaviour of the underlying time series. Neither the GBM-generated training data nor the real training data consisting of the 25 multi-asset futures contracts seems to correctly mirror these characteristics. Moreover, there are strong hints provided by our results that over-fitting is still present to some extend.

Based on these insights, here is the road-map for future research:

\begin{enumerate}[nosep]
    \item Modify the reward functions towards a risk-adjusted framework. We use a very simple one without considering transaction costs and volatility. To refine the model, investigate more complicated reward functions to better simulate real trading practice.
    \item Expand state spaces and action spaces. Expand the state spaces by adding other input features to improve the accuracy of simulating the real financial market environment and to minimize over-fitting. In addition, continuous action spaces can be introduced so that multiple models like PPO, A2C are applicable and tradings are more flexible.
    \item Extend the state space by including a broader diversity of international equity index futures contracts as training data.
    \item "Even more of the same": Generate more artificial training data sets to a) overcome over-fitting and to b) account for a the highly-divers volatility regimes.
    \item Generate/simulate artificial training data for a variety of different markets and asset classes. This, however, raises the difficult challenge to, for each simulated scenario generated, retain full plausibility with regard to volatility, correlation, and other macro parameters.
    \item Utilize customized neural networks. To make the model more suitable for financial time series data, use customized neural networks such as CNNs, RNNs, and LSTMs to approximate the Q-function and tune the parameters.
\end{enumerate}

\bibliographystyle{unsrt}  
\bibliography{references}

\clearpage

\appendix

\section{Literature Appendix}
\label{appendix literature}

\begin{enumerate}[nosep]
    \item \textbf{Portfolio Management using Reinforcement Learning} \cite{jin2016portfolio}
    \newline 
    \newline This project, undertaken, by Olivier Jin and Hamza El-Saawy from Stanford University uses DQN to train a neural network to manage a portfolio containing one high-volatility stock and one low-volatility stock. They take stock histories for both stocks over a set number of days (either 2, 7, or 30), the inventory of each stock, the total portfolio value, and the left-over cash to represent states. 
    \newline
    \newline The action space is a set of discrete percentages of the portfolio's total value each stock composes.  The agent chooses an action $a_t$ from $[-0.25, -0.1, 0.05, 0, 0.05, 0.1, 0.25]$ to represent a selling of ($a_t\times$ portfolio's total value) of low-volatility stock and a buying of the corresponding amount of the high-volatility stock (and vice versa for $a_t < 0$). A fixed transaction cost of \$0.001 per transaction is also introduced. 
    \newline
    \newline In terms of reward functions, they modify the portfolio return by penalizing volatility: $P_t = R_t - \lambda std(R_t)$; $\forall t \in[1, \mathrm{T}]$, where $R_t = v_t - v_{t-1}$, and $v_t$ is portfolio return at time $t$. To balance the exploration \& exploitation and decorrelate the time series data, an $\epsilon$ - greedy exploration strategy and experience replay are adopted respectively. Instead of using a single neural network to approximate the Q function, the weights of the target network are updated with the main network's weights. The performance of models is compared against two benchmarks. 
    \newline
    \newline The results show overall that the DQN model outperforms benchmarks with a higher Sharpe ratio and significantly less variance. Their work has indicated that RL is feasible in Portfolio Management.
    \newline
    \item \textbf{Financial Trading as a Game: A Deep Reinforcement Learning Approach} \cite{huang2018financial}
    \newline
    \newline This paper by Chien-Yi Huan talks about the effectiveness of applying deep reinforcement learning algorithms to the financial trading domain. 
    \newline 
    \newline The author uses the following steps to deal with this problem:
    \begin{enumerate}[nosep] 
        \item Proposes a Markov decision process (MDP) model for general signal-based financial trading tasks
        \item Modifies the existing deep recurrent Q-network algorithm to be suitable for the financial trading task and also uses a substantially smaller replay memory and sampling a longer sequence for training
        \item Discovers workable hyperparameters for the DRQN algorithm and develops a novel action augmentation technique to mitigate the need for random exploration in the financial trading environment
        \item Achieves positive return on 12 different currency pairs including major and cross pairs under transaction costs
        \item And lastly discovers a counter-intuitive fact that a slightly increased spread leads to the better overall performance
    \end{enumerate}
    The state space is defined in three parts: time feature, market feature, and position feature.
    \newline
    \newline The action space is discrete and is a simple action set of three values $\{-1, 0, 1\}$. Position reversal is allowed and the reward function is the portfolio log-returns at each time step.
    \newline
    \item \textbf{A Deep Reinforcement Learning Framework for the Financial Portfolio Management Problem} \cite{jiang2017deep}
    \newline
    \newline In this paper the authors, Zhengyao Jiang, Dixing Xu and Jinjun Liang, present a financial-model-free Reinforcement Learning framework to provide a deep machine learning solution to the portfolio management problem by using three different underlying networks, a Convolutional Neural Network (CNN), a basic Recurrent Neural Network (RNN), and a Long Short-Term Memory (LSTM). 
    \newline 
    \newline Based on the assumptions of zero market impact and zero slippage, they examine the models in three back-test experiments with a trading period of 30 minutes in a cryptocurrency market along with some recently reviewed or published portfolio-selection strategies. All three instances of the framework monopolize the top three positions in all experiments, outdistancing other compared trading algorithms. Although with a high commission rate of 0.25\% in the backtests, the framework can achieve at least 4-fold returns in 50 days.
    \newline
    \item \textbf{Practical Deep Reinforcement Learning Approach for Stock Trading} \cite{xiong2018practical}
    \newline 
    \newline This paper by Zhuoran Xiong, Xiao-Yang Liu, Shan Zhong, Hongyang Yang, and Anwar Walid explores the potential of deep reinforcement learning to optimize stock trading strategy and thus maximize investment return. 30 stocks are selected as trading stocks and their daily prices are used as the training and trading market environment. They also explore the potential of training Deep Deterministic Policy Gradient (DDPG) agents to learn stock trading strategy. 
    \newline 
    \newline The approach in this paper is they first build the environment by setting 30 stocks data as a vector of daily stock prices over which the DDPG agent is trained. DDPG is an improved version of Deterministic Policy Gradient (DPG) algorithm and it combines the frameworks of both Q-learning and policy gradient. Compared with the DPG algorithm, DDPG uses neural networks as function approximators. To update the learning rate and the number of episodes, the agent is validated on validation data.
    \newline 
    \newline The agent's performance is then evaluated and compared with Dow Jones Industrial Average and the traditional min-variance portfolio allocation strategy. Results show that the proposed deep reinforcement learning approach is shown to outperform the two baselines in terms of both the Sharpe ratio and cumulative returns. The comparison of Sharpe ratios indicates the method is more robust than the others in balancing risk and return.
    \newline
    \item \textbf{Optimal asset allocation using adaptive dynamic programming} \cite{neuneier1996optimal}
    \newline 
    \newline In this paper by Ralph Neuneier, asset allocation is formalized as a Markovian Decision Problem which can be optimized by applying dynamic programming or reinforcement learning based algorithms. Using an artificial exchange rate, the asset allocation strategy optimized with reinforcement learning (Q-Learning) is shown to be equivalent to a policy computed by dynamic programming. 
    \newline 
    \newline Initially the modeling phase and the search for an optimal portfolio are combined and embedded in the framework of Markovian Decision Problems, MDP. If the discrete state space is small and if an accurate model of the system is available, MDP can be solved by conventional Dynamic Programming, DP. On the other extreme, reinforcement learning methods, e.g. Q-Learning, QL, can be applied to problems with large state spaces and with no appropriate model available. 
    \newline 
    \newline The approach is then tested on the task to invest liquid capital in the German stock market. Here, neural networks are used as value function approximators. The resulting asset allocation strategy is superior to a heuristic benchmark policy. 
\end{enumerate}

\clearpage

\section{Data Appendix}
\label{appendix data}

\begin{table}
\caption{Description of 25 continuous futures}
\centering
\begin{tabular}{llll}
\hline
CLASS & EXCHANGE & SYMBOL & NAME \\ \hline
Equity Indexes & CME & ES & CME S\&P 500 Index E-Mini \\
 & CME & MD & CME S\&P 400 Midcap Index \\
 & CME & NK & CME Nikkei 225 \\
 & CME & NQ & CME NASDAQ 100 Index Mini \\
 & CME & SP & CME S\&P 500 Index \\ \hline
Fixed Incomes & CME & FV & CBOT 5-year US Treasury Note \\
 & CME & TY & CBOT 10-year US Treasury Note \\
 & CME & US & CBOT 30-year US Treasury Bond \\ \hline
Forex & CME & AD & CME Australian Dollar AUD \\
 & CME & BP & CME British Pound GBP \\
 & CME & CD & CME Canadian Dollar CAD \\
 & CME & EC & CME Euro FX \\
 & CME & JY & CME Japanese Yen JPY \\
 & CME & SF & CME Swiss Franc CHF \\
 & ICE & DX & ICE US Dollar Index \\ \hline
Commodities & CME & C & CBOT Corn \\
 & CME & CL & NYMEX WTI Crude Oil \\
 & CME & GC & NYMEX Gold \\
 & CME & HO & NYMEX Heating Oil \\
 & CME & LC & CME Live Cattle \\
 & CME & NG & NYMEX Natural Gas \\
 & CME & S & CBOT Soybeans \\
 & CME & SI & NYMEX Silver \\
 & CME & W & CBOT Wheat \\
 & ICE & SB & ICE Sugar No. 11 \\ \hline
 \label{tab:contratcs}
\end{tabular}
\end{table}

\clearpage

\begin{figure}[ht]
     \centering
     \begin{subfigure}[b]{0.48\textwidth}
         \centering
         \includegraphics[width=\textwidth]{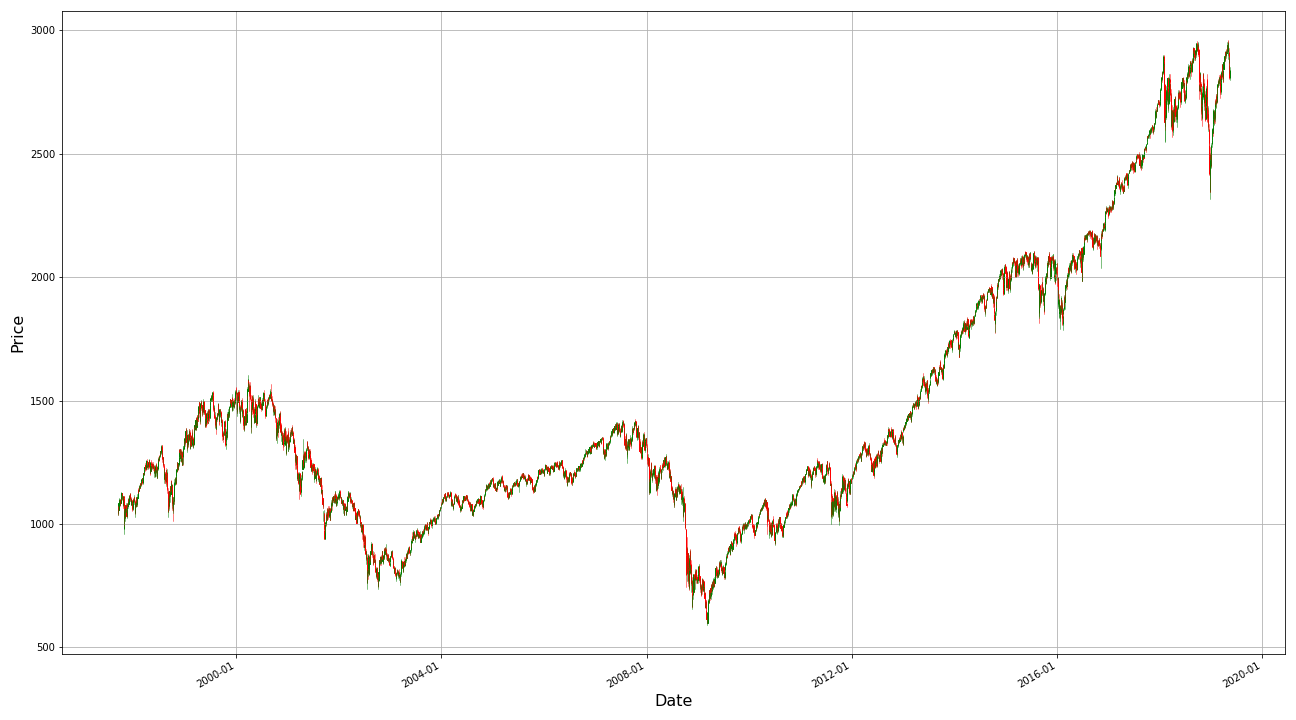}
     \end{subfigure}
     \hfill
     \begin{subfigure}[b]{0.48\textwidth}
         \centering
         \includegraphics[width=\textwidth]{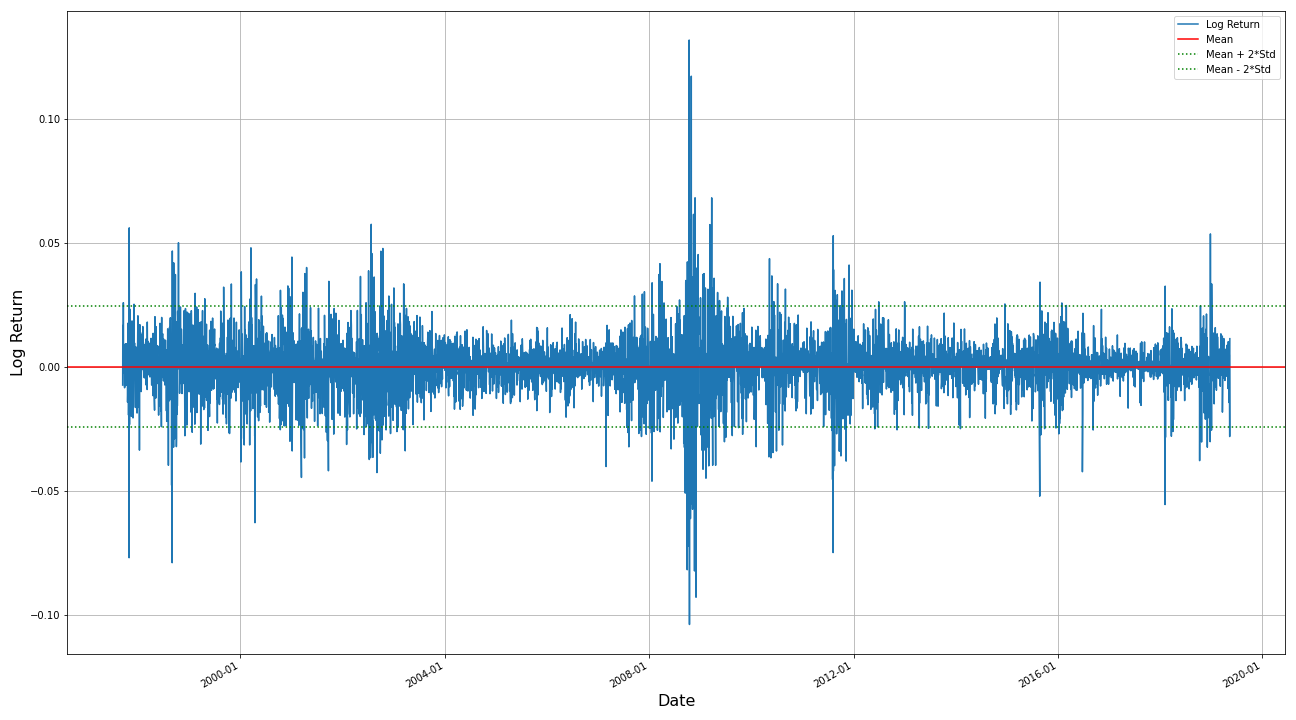}
     \end{subfigure}
    \caption{Daily candlestick chart (left) and log return (right) of E-mini S\&P 500 continuous future}
\end{figure}

\begin{figure}[ht]
     \centering
     \begin{subfigure}[b]{0.48\textwidth}
         \centering
         \includegraphics[width=\textwidth]{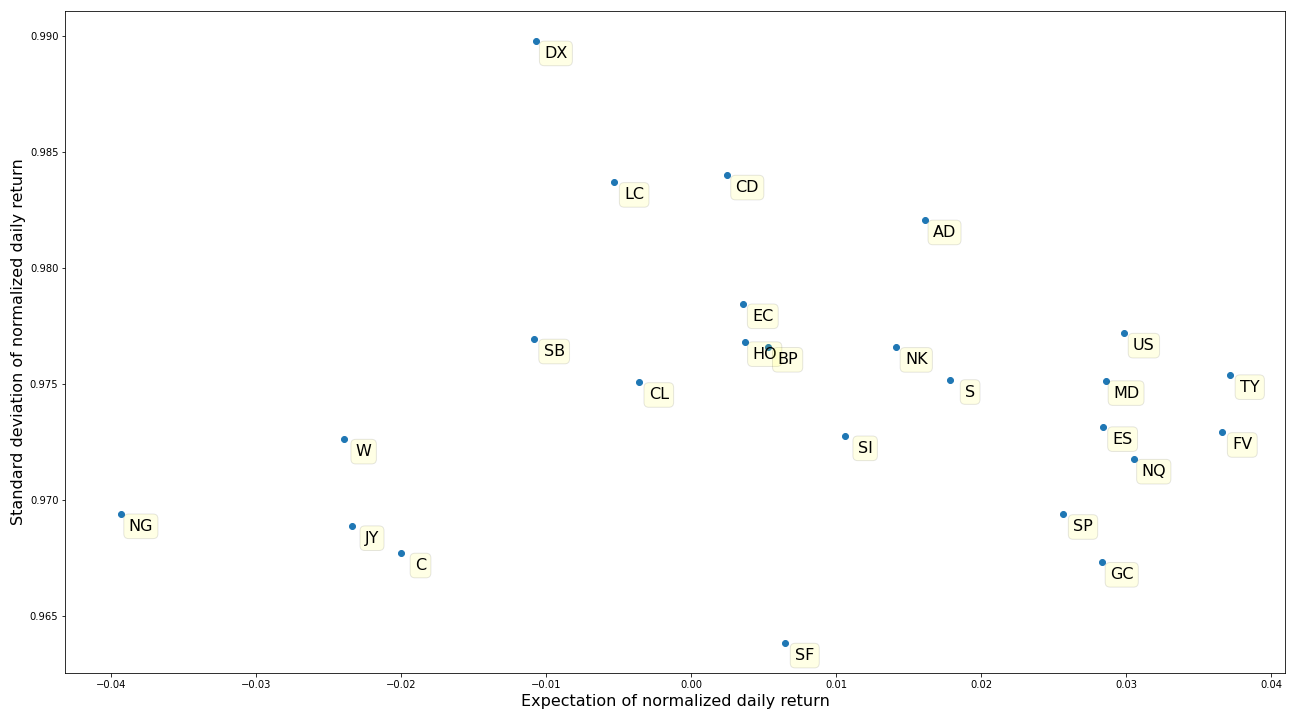}
     \end{subfigure}
     \hfill
     \begin{subfigure}[b]{0.48\textwidth}
         \centering
         \includegraphics[width=\textwidth]{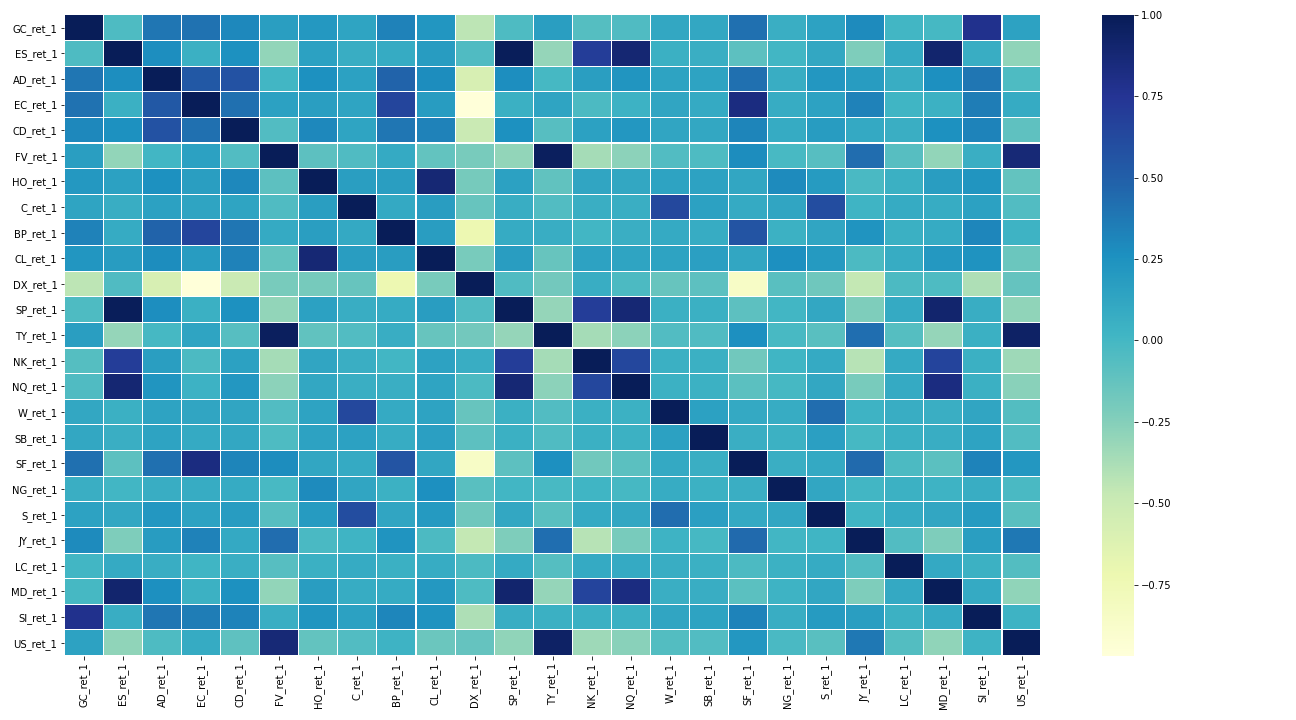}
     \end{subfigure}
    \caption{Annualised expectation and standard deviation of returns of multi-asset futures (left), and correlation between them (right)}
\end{figure}

\begin{figure}[ht]
     \centering
     \begin{subfigure}[b]{0.48\textwidth}
         \centering
         \includegraphics[width=\textwidth]{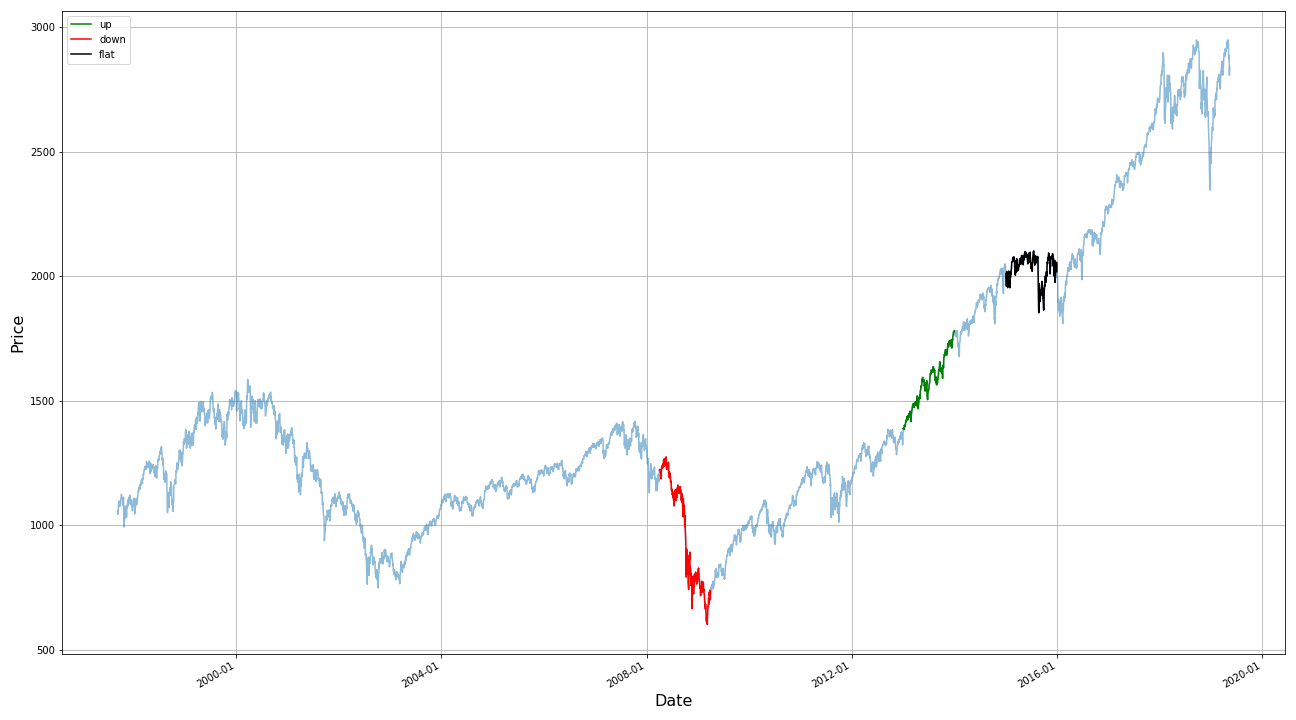}
     \end{subfigure}
     \hfill
     \begin{subfigure}[b]{0.48\textwidth}
         \centering
         \includegraphics[width=\textwidth]{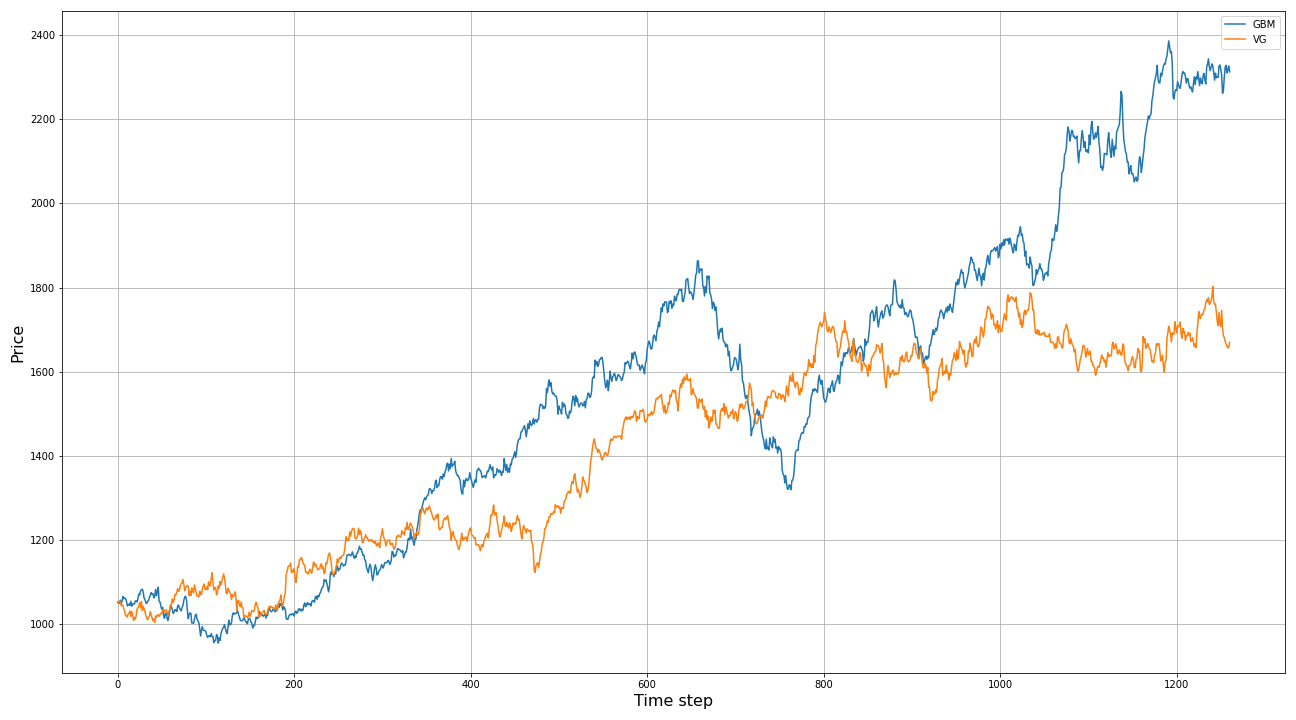}
     \end{subfigure}
    \caption{Selected real data for parameter estimation (left), and examples of GBM and VG processes (right)}
    \small\textsuperscript{down-trend: 04/01/2008 - 04/01/2009, up-trend: 01/01/2013 - 01/01/2014, no-trend: 01/01/2015 - 01/01/2016}
    \label{fig:regimes}
\end{figure}

\clearpage

\section{Model Appendix}
\label{appendix model}

We have two models with the same architecture for Q function and target Q function respectively. Each model contains two neural networks as approximators of the action-value function and the state-value function. And each network has two hidden layers with 64 units, with $\tanh$ as activation function, and with layer normalization. 

\begin{table}
\centering
\caption{Shape of parameters in neural networks}
\begin{tabular}{l|l|lll}
\hline
eps &  &  &  & (1,) \\ \hline
model & action\_value & fully\_connected & weights & (180,64) \\
 &  &  & biases & (64,) \\
 &  & LayerNorm & beta & (64,) \\
 &  &  & gamma & (64,) \\
 &  & fully\_connected\_1 & weights & (64,64) \\
 &  &  & biases & (64,) \\
 &  & LayerNorm\_1 & beta & (64,) \\
 &  &  & gamma & (64,) \\
 &  & fully\_connected\_2 & weights & (64,3) \\
 &  &  & biases & (3,) \\
 & state\_value & fully\_connected & weights & (180,64) \\
 &  &  & biases & (64,) \\
 &  & LayerNorm & beta & (64,) \\
 &  &  & gamma & (64,) \\
 &  & fully\_connected\_1 & weights & (64,64) \\
 &  &  & biases & (64,) \\
 &  & LayerNorm\_1 & beta & (64,) \\
 &  &  & gamma & (64,) \\
 &  & fully\_connected\_2 & weights & (64,1) \\
 &  &  & biases & (1,) \\\hline
target\_q\_func: model & action\_value & fully\_connected & weights & (180,64) \\
 &  &  & biases & (64,) \\
 &  & LayerNorm & beta & (64,) \\
 &  &  & gamma & (64,) \\
 &  & fully\_connected\_1 & weights & (64,64) \\
 &  &  & biases & (64,) \\
 &  & LayerNorm\_1 & beta & (64,) \\
 &  &  & gamma & (64,) \\
 &  & fully\_connected\_2 & weights & (64,3) \\
 &  &  & biases & (3,) \\
 & state\_value & fully\_connected & weights & (180,64) \\
 &  &  & biases & (64,) \\
 &  & LayerNorm & beta & (64,) \\
 &  &  & gamma & (64,) \\
 &  & fully\_connected\_1 & weights & (64,64) \\
 &  &  & biases & (64,) \\
 &  & LayerNorm\_1 & beta & (64,) \\
 &  &  & gamma & (64,) \\
 &  & fully\_connected\_2 & weights & (64,1) \\
 &  &  & biases & (1,)\\\hline
\end{tabular}
\label{tab:modparas}
\end{table}

\clearpage

\section{Result Appendix}
\label{appendix result}

\begin{figure}[ht]
     \centering
     \begin{subfigure}[b]{0.33\textwidth}
         \centering
         \includegraphics[width=\textwidth]{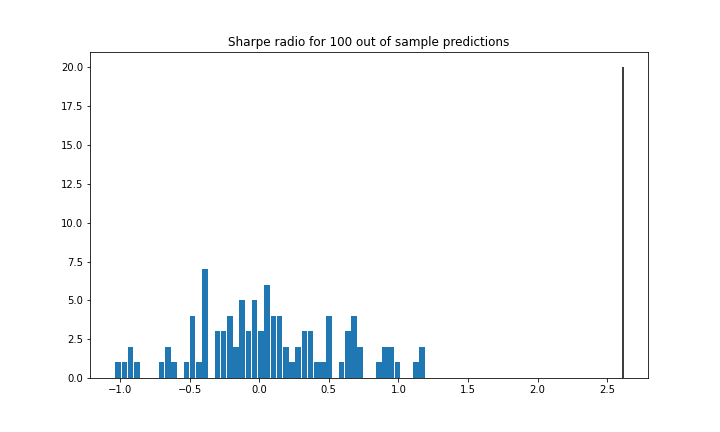}
         \caption{GBM, up-trend, 1 path}
     \end{subfigure}
     \hfill
     \begin{subfigure}[b]{0.33\textwidth}
         \centering
         \includegraphics[width=\textwidth]{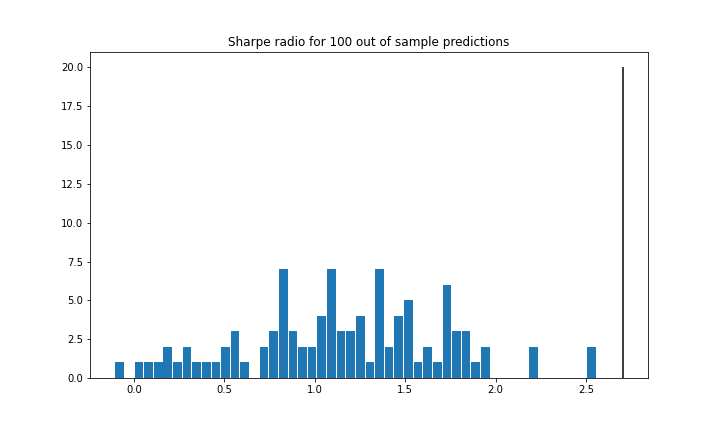}
         \caption{GBM, up-trend, 50 paths}
     \end{subfigure}
     \hfill
     \begin{subfigure}[b]{0.33\textwidth}
         \centering
         \includegraphics[width=\textwidth]{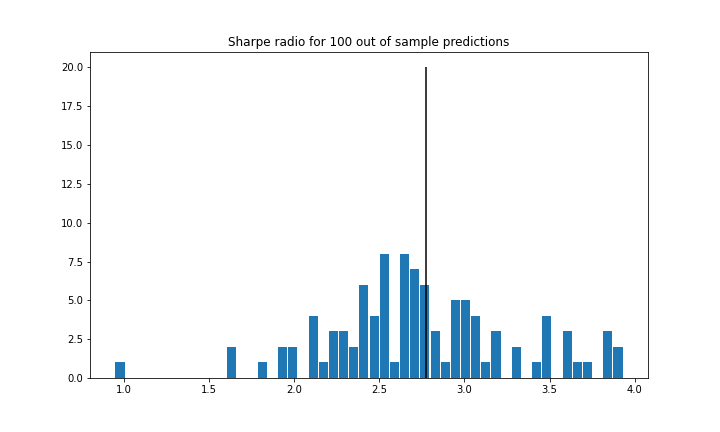}
         \caption{GBM, up-trend, 90 paths}
     \end{subfigure}
     \newline
     \begin{subfigure}[b]{0.33\textwidth}
         \centering
         \includegraphics[width=\textwidth]{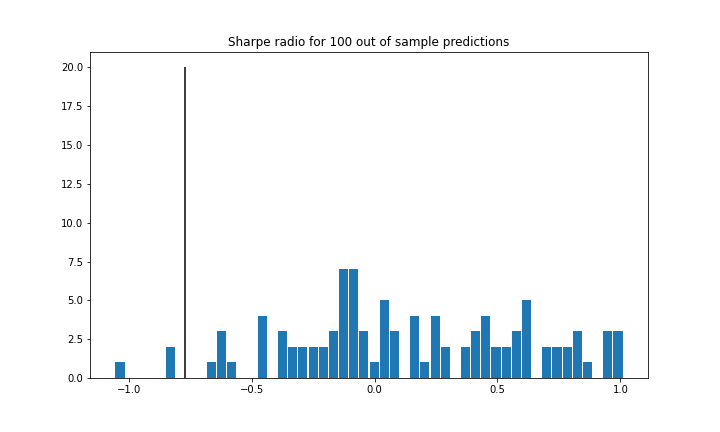}
         \caption{GBM, down-trend, 1 path}
     \end{subfigure}
     \hfill
     \begin{subfigure}[b]{0.33\textwidth}
         \centering
         \includegraphics[width=\textwidth]{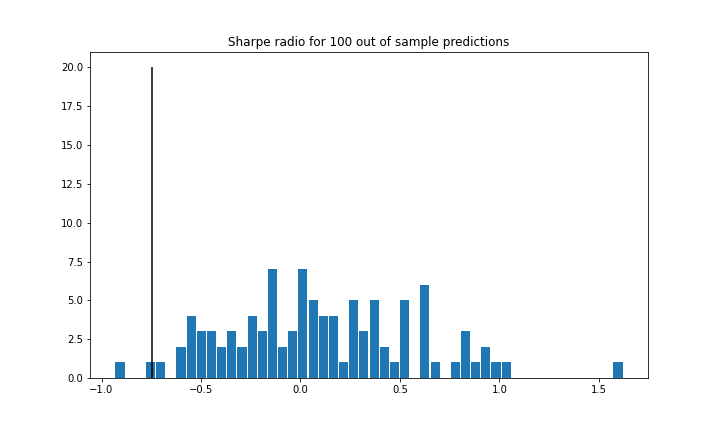}
         \caption{GBM, down-trend, 50 paths}
     \end{subfigure}
     \hfill
     \begin{subfigure}[b]{0.33\textwidth}
         \centering
         \includegraphics[width=\textwidth]{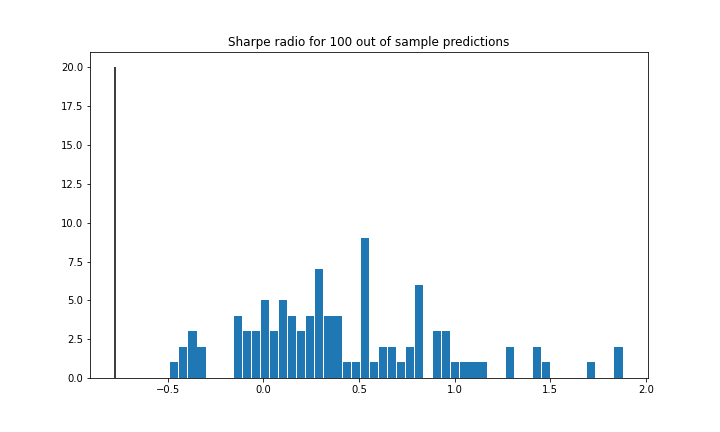}
         \caption{GBM, down-trend, 90 paths}
     \end{subfigure}
     \newline
     \begin{subfigure}[b]{0.33\textwidth}
         \centering
         \includegraphics[width=\textwidth]{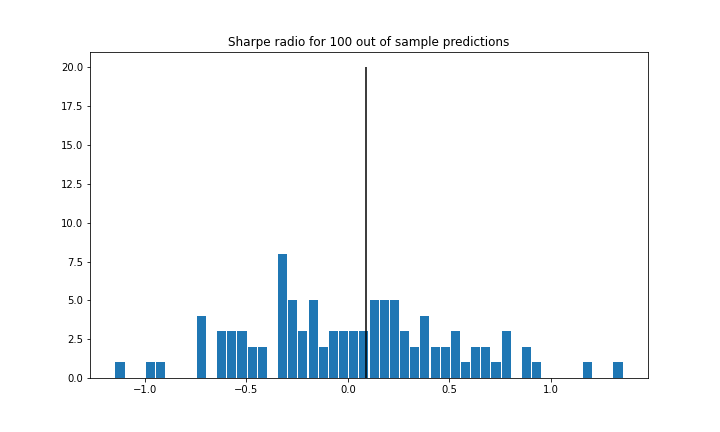}
         \caption{GBM, no-trend, 1 path}
     \end{subfigure}
     \hfill
     \begin{subfigure}[b]{0.33\textwidth}
         \centering
         \includegraphics[width=\textwidth]{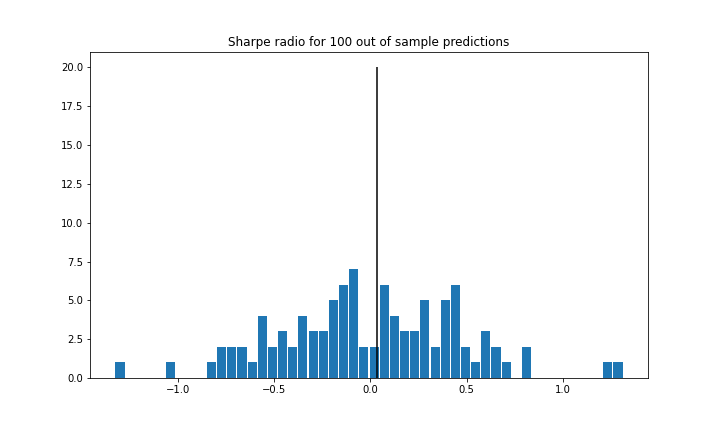}
         \caption{GBM, no-trend, 50 paths}
     \end{subfigure}
     \hfill
     \begin{subfigure}[b]{0.33\textwidth}
         \centering
         \includegraphics[width=\textwidth]{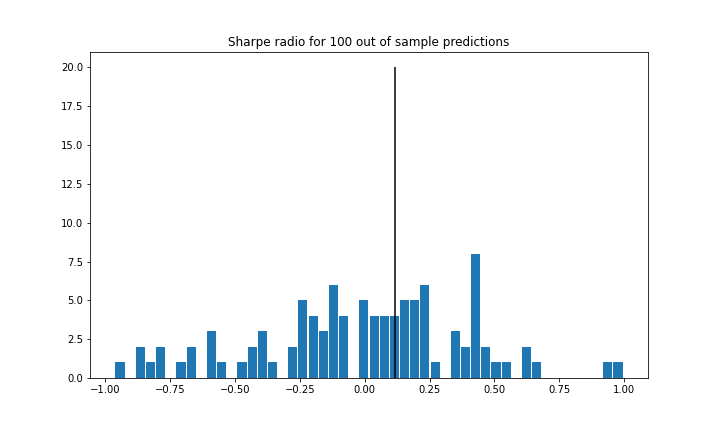}
         \caption{GBM, no-trend, 90 paths}
     \end{subfigure}
     \newline
     \begin{subfigure}[b]{0.33\textwidth}
         \centering
         \includegraphics[width=\textwidth]{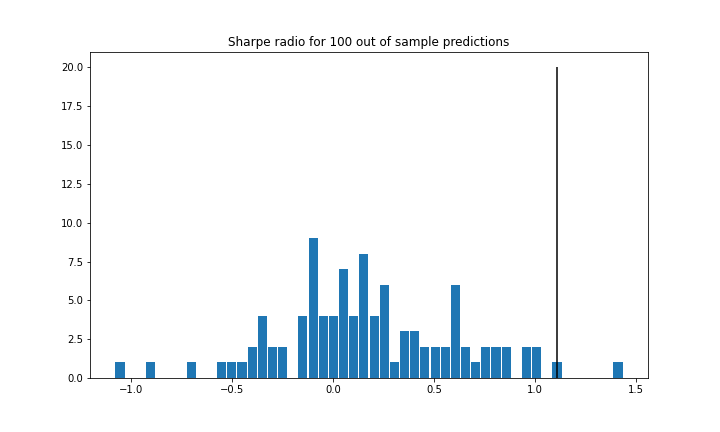}
         \caption{GBM, switch-trend, 1 path}
     \end{subfigure}
     \hfill
     \begin{subfigure}[b]{0.33\textwidth}
         \centering
         \includegraphics[width=\textwidth]{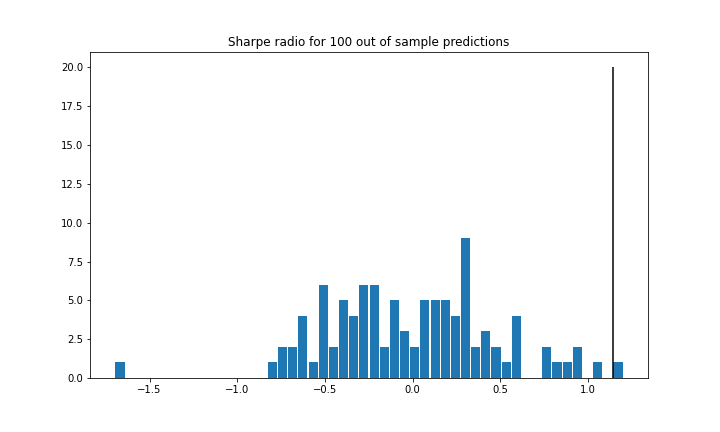}
         \caption{GBM, switch-trend, 50 paths}
     \end{subfigure}
     \hfill
     \begin{subfigure}[b]{0.33\textwidth}
         \centering
         \includegraphics[width=\textwidth]{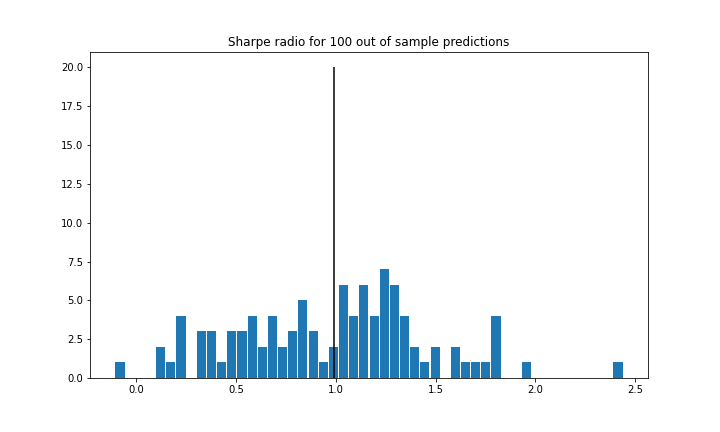}
         \caption{GBM, switch-trend, 90 paths}
     \end{subfigure}
    \caption{Histograms of annualized Sharpe Ratio, using \textbf{GBM} as simulator, for up (1st row), down (2nd row), no (3rd row), and switch (4th row) \textbf{trend}, and for 1 (1st column), 50 (2nd column), and 90 (3rd column) \textbf{paths} of the process for training. The vertical line represents the average of the annualized Sharpe Ratio of the benchmark}
\label{fig:gbm}
\end{figure}

\begin{figure}[ht]
     \centering
     \begin{subfigure}[b]{0.33\textwidth}
         \centering
         \includegraphics[width=\textwidth]{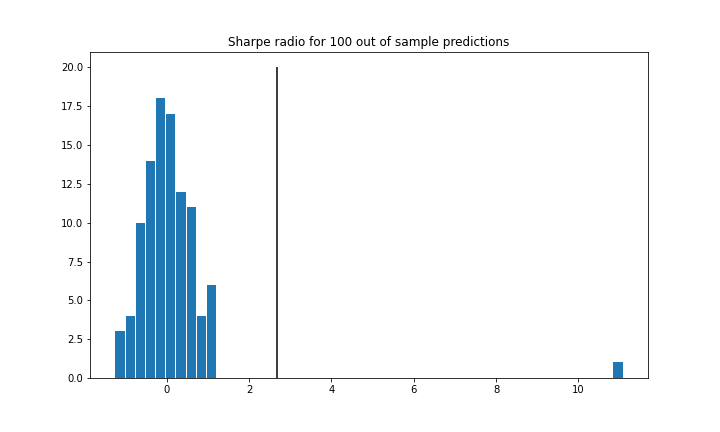}
         \caption{VG, up-trend, 1 path}
     \end{subfigure}
     \hfill
     \begin{subfigure}[b]{0.33\textwidth}
         \centering
         \includegraphics[width=\textwidth]{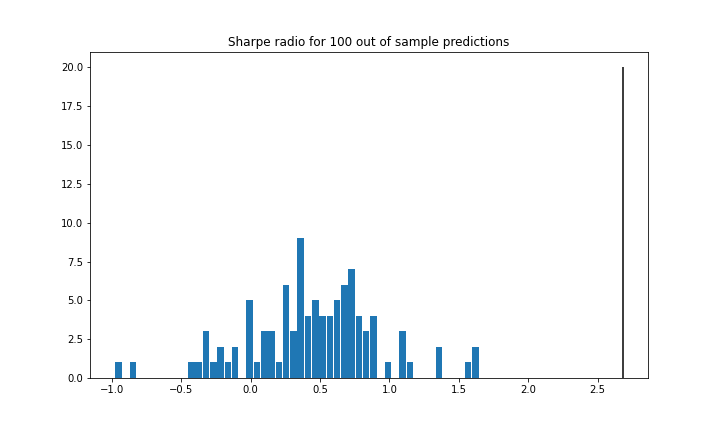}
         \caption{VG, up-trend, 50 paths}
     \end{subfigure}
     \hfill
     \begin{subfigure}[b]{0.33\textwidth}
         \centering
         \includegraphics[width=\textwidth]{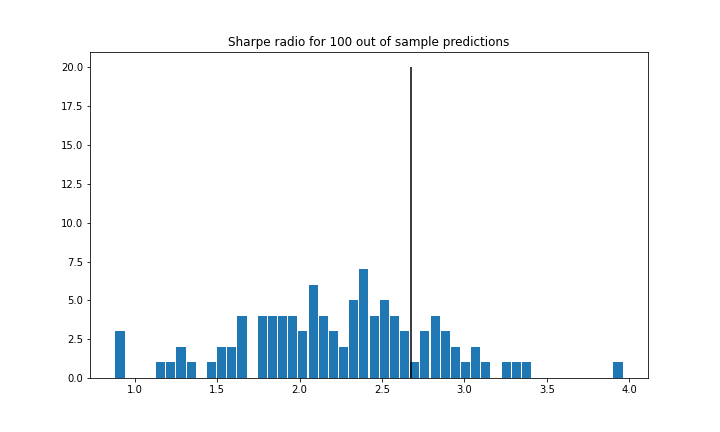}
         \caption{VG, up-trend, 90 paths}
     \end{subfigure}
     \newline
     \begin{subfigure}[b]{0.33\textwidth}
         \centering
         \includegraphics[width=\textwidth]{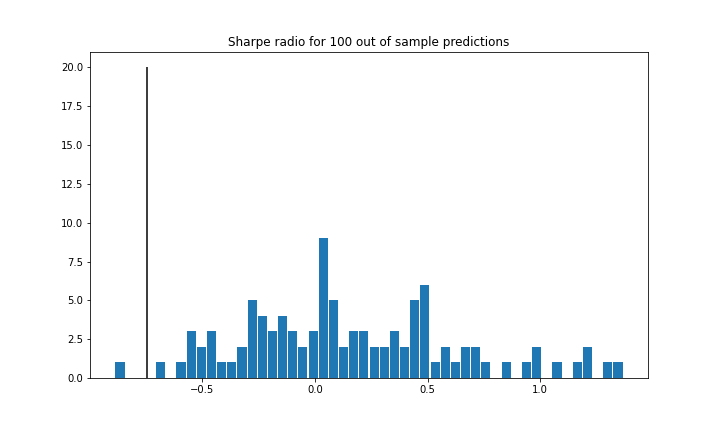}
         \caption{VG, down-trend, 1 path}
     \end{subfigure}
     \hfill
     \begin{subfigure}[b]{0.33\textwidth}
         \centering
         \includegraphics[width=\textwidth]{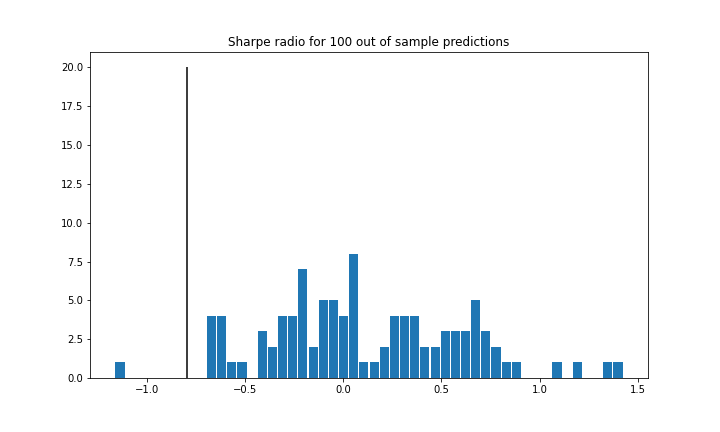}
         \caption{VG, down-trend, 50 paths}
     \end{subfigure}
     \hfill
     \begin{subfigure}[b]{0.33\textwidth}
         \centering
         \includegraphics[width=\textwidth]{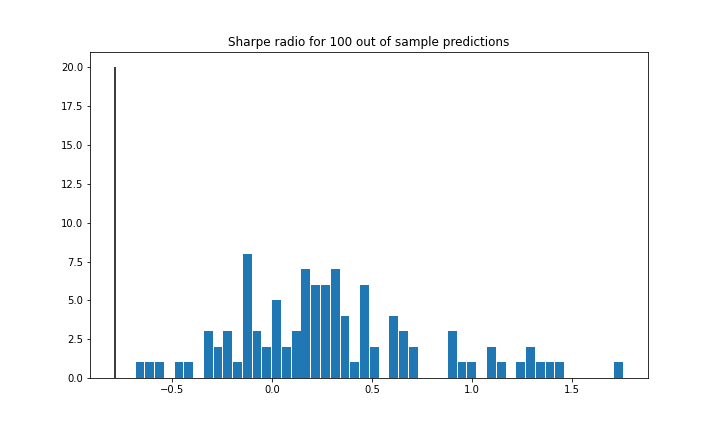}
         \caption{VG, down-trend, 90 paths}
     \end{subfigure}
     \newline
     \begin{subfigure}[b]{0.33\textwidth}
         \centering
         \includegraphics[width=\textwidth]{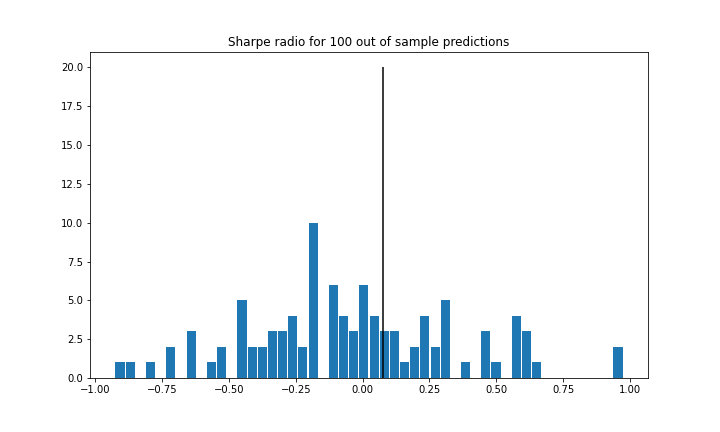}
         \caption{VG, no-trend, 1 path}
     \end{subfigure}
     \hfill
     \begin{subfigure}[b]{0.33\textwidth}
         \centering
         \includegraphics[width=\textwidth]{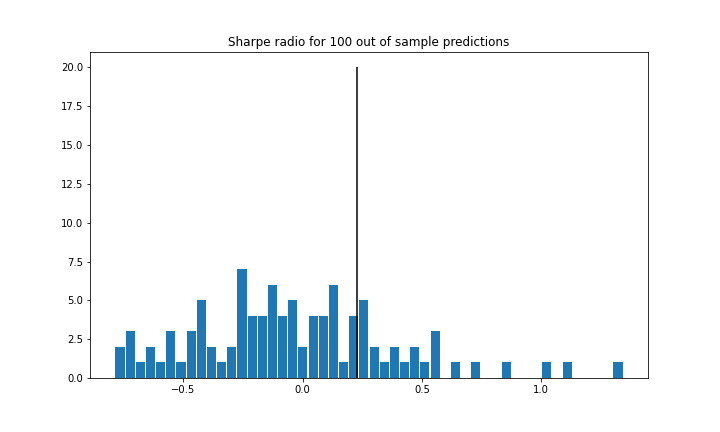}
         \caption{VG, no-trend, 50 paths}
     \end{subfigure}
     \hfill
     \begin{subfigure}[b]{0.33\textwidth}
         \centering
         \includegraphics[width=\textwidth]{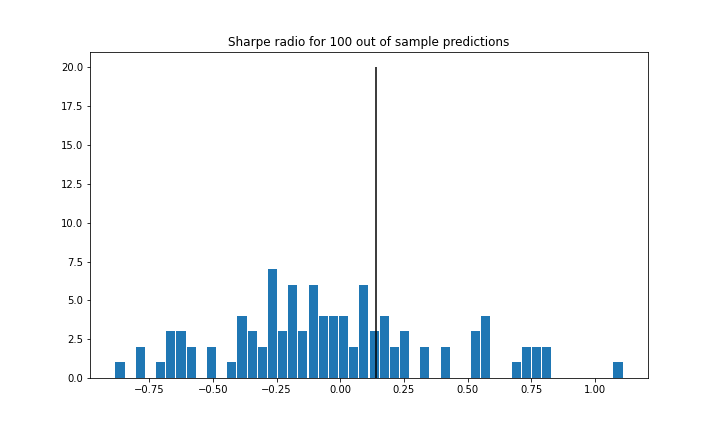}
         \caption{VG, no-trend, 90 paths}
     \end{subfigure}
     \newline
     \begin{subfigure}[b]{0.33\textwidth}
         \centering
         \includegraphics[width=\textwidth]{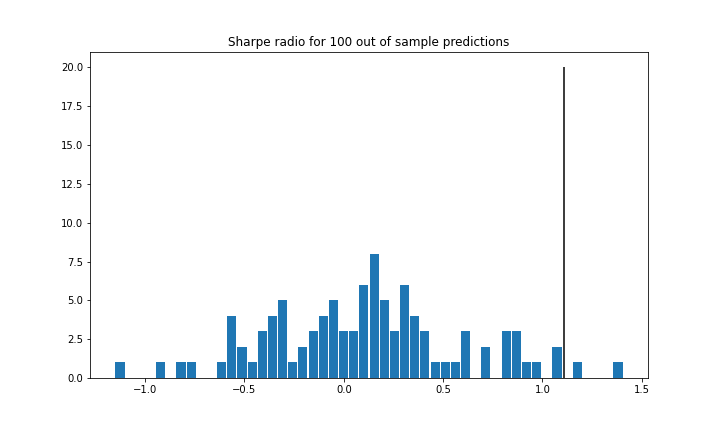}
         \caption{VG, switch-trend, 1 path}
     \end{subfigure}
     \hfill
     \begin{subfigure}[b]{0.33\textwidth}
         \centering
         \includegraphics[width=\textwidth]{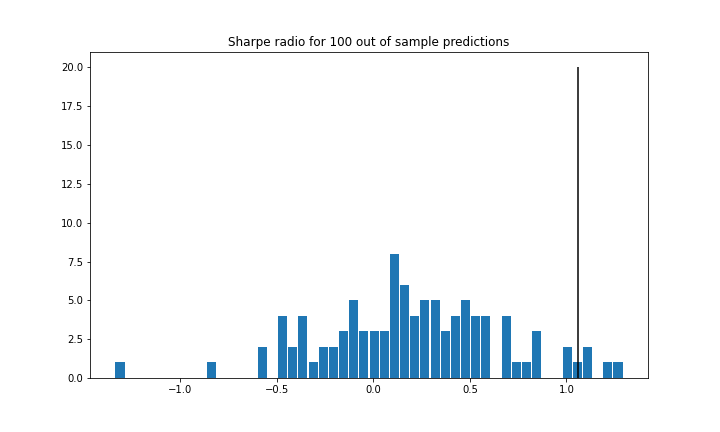}
         \caption{VG, switch-trend, 50 paths}
     \end{subfigure}
     \hfill
     \begin{subfigure}[b]{0.33\textwidth}
         \centering
         \includegraphics[width=\textwidth]{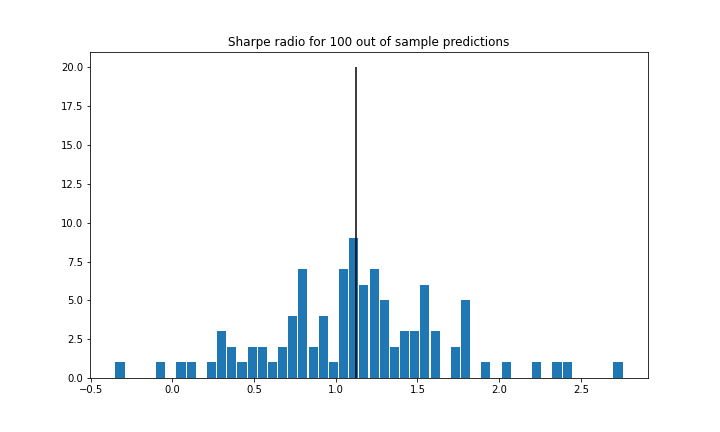}
         \caption{VG, switch-trend, 90 paths}
     \end{subfigure}
    \caption{Histograms of annualized Sharpe Ratio, using \textbf{VG} as simulator, for up (1st row), down (2nd row), no (3rd row), and switch (4th row) \textbf{trend}, and for 1 (1st column), 50 (2nd column), and 90 (3rd column) \textbf{paths} of the process for training. The vertical line represents the average of the annualized Sharpe Ratio of the benchmark.}
\label{fig:vg}
\end{figure}

\end{document}